\documentclass[%
 aip,
 amsmath,amssymb,
 reprint,%
]{revtex4-1}
\usepackage{graphicx}% Include figure files
\usepackage{dcolumn}% Align table columns on decimal point
\usepackage{bm}% bold math
\usepackage[utf8]{inputenc}
\usepackage[T1]{fontenc}
\usepackage{mathptmx}
\usepackage{etoolbox}

\usepackage{xcolor}

\makeatletter
\def\@email#1#2{%
 \endgroup
 \patchcmd{\titleblock@produce}
  {\frontmatter@RRAPformat}
  {\frontmatter@RRAPformat{\produce@RRAP{*#1\href{mailto:#2}{#2}}}\frontmatter@RRAPformat}
  {}{}
}%
\makeatother

\graphicspath{{figures/}}

\begin{document}

\preprint{AIP/123-QED}

\title[Role of coupling delay in oscillatory activity in autonomous networks of excitable neurons with dissipation]{Role of coupling delay in oscillatory activity in autonomous networks of excitable neurons with dissipation}
%\title[Impact of time delay on the dynamics of coupled FitzHugh--Nagumo neurons with dissipation]{Impact of time delay on the dynamics of coupled FitzHugh--Nagumo neurons with dissipation}
% Force line breaks with \\
\author{A. V. Bukh}
\affiliation{Institute of Physics, Saratov State University, 83 Astrakhanskaya Street, Saratov 410012, Russia.}%Lines break automatically or can be forced with \\
\author{I. A. Shepelev}
\affiliation{Institute of Physics, Saratov State University, 83 Astrakhanskaya Street, Saratov 410012, Russia.}%Lines break automatically or can be forced with \\
\affiliation{Almetyevsk State Petroleum Institute, 2 Lenina Street, Almetyevsk 423458, Tatarstan, Russia}%Lines break automatically or can be forced with \\
\author{E. M. Elizarov}
\affiliation{Institute of Physics, Saratov State University, 83 Astrakhanskaya Street, Saratov 410012, Russia.}%Lines break automatically or can be forced with \\
\author{S. S. Muni}
\affiliation{Department of Physical Sciences, Indian Institute of Science Education and Research Kolkata, Campus Road, Mohanpur, West Bengal 741246, India.}%Lines break automatically or can be forced with \\
\author{E. Sch\"oll}
\affiliation{Institut f\"ur Theoretische Physik, Technische
Universit\"at Berlin, Berlin, Germany; Bernstein Center for Computational Neuroscience Berlin, Berlin, Germany; Potsdam Institute for Climate Impact Research, Potsdam, Germany.}%Lines break automatically or can be forced with \\
\author{G. I. Strelkova}%
%\email{strelkovagi@sgu.ru}
\affiliation{Institute of Physics, Saratov State University, 83 Astrakhanskaya Street, Saratov 410012, Russia.}%Lines break automatically or can be forced with \\

\date{\today}% It is always \today, today,

\begin{abstract}
We study numerically the effects of time delay in networks of delay-coupled excitable FitzHugh--Nagumo systems with dissipation. Generation of periodic self-sustained oscillations and its threshold are analyzed depending on the dissipation of a single neuron, the delay time  and random initial conditions.  The peculiarities of spatiotemporal dynamics of time-delayed bidirectional ring-structured FitzHugh--Nagumo neuronal systems are investigated in cases of local and nonlocal coupling topology between the nodes, and a first-order nonequilibrium phase transition to synchrony is established. It is shown that the emergence of oscillatory activity in delay-coupled FitzHugh--Nagumo neurons is observed for smaller values of the coupling strength as the dissipation parameter decreases. This can provide the possibility of controlling the spatio-temporal behavior of the considered neuronal networks. The observed effects are quantified by plotting distributions of the maximal Lyapunov exponent and the global order parameter in terms of delay and coupling strength.
\end{abstract}

\maketitle

\begin{quotation}
Excitability is a common property of many physical and biological systems. Systems consisting of coupled excitable elements have been widely studied as models for many natural phenomena, such as the communication between neurons in the
brain. Since the work of Hodgkin and Huxley\cite{Hodgkin1952vw}, and the development of the basic mathematical model by FitzHugh\cite{fitzhugh1961impulses}, and Nagumo et al.\cite{Nagumo1962} the reported research on the subject has grown enormously.
Delays are inherent in neuronal networks due to finite conduction velocities and synaptic transmission.  Recently, many researchers have investigated the effects of time delays in neuronal networks and found many delay-induced phenomena. In particular, it has been established that time-delayed coupling can be used as a powerful tool for stabilizing various complex spatiotemporal patterns and for controlling different types of synchronization in one- and multilayer networks.  
In the present work we consider the FitzHugh--Nagumo oscillator which represents a paradigmatic model for neuronal excitability. The considered two-variable system of equations includes an additional parameter which takes into account the dissipation of a neuron and thus differs from the simplified FitzHugh--Nagumo model.
Starting from a pair of delay-coupled FitzHugh--Nagumo oscillators and proceeding
via simple ring neuronal networks with delayed coupling, we analyze how the spatiotemporal dynamics of the networks depends on the parameter of dissipation, the delay time, the coupling parameters and randomly distributed initial conditions. 
We provide linear stability analyses and plot  distributions of the maximal Lyapunov exponent and the global order parameter depending 
on the delay time and the coupling strength. Furthermore, we establish nonequilibrium phase transitions from multiple clusters to full synchronization. The presented results are compared with the previously obtained findings for networks of simplified FitzHugh--Nagumo models.
\end{quotation}

\section{\label{sec:intro}Introduction}

Excitable systems play important roles in understanding and modelling various natural phenomena, such as, for example,  the transmission of impulses between neurons in the brain, the cardiac arrhythmia, the appearance of organized structures in the cortex of egg cells, etc.\cite{Keener1998, Fall2002, Ermentrout2010}
% modelling various phenomena (effects) occurring in physical and biological systems.
Generation of a single spike in the electrical potential across the neuron
membrane is a typical example of excitable behavior. Such excitable units usually appear as constitutive elements of complex systems, and can transmit excitations between them. The phenomenological FitzHugh--Nagumo set of equations, which is a two-dimensional simplification of the four-variable Hodgkin-Huxley model, has been proven to be a successful model for the description of the spike generation of the neuronal axon\cite{fitzhugh1961impulses, Nagumo1962}. A biological neuron is a dissipative object (unit) in which a single spike activity decays rather quickly due to high electrical resistance. Accordingly, the higher the dissipation in the neuron, the greater the energy required to excite the spike. The dissipative nature of spike generation in neurons was described in Refs.\cite{Appali2012,Lindner2022} for some neuronal models. The presence of dissipation in the FitzHugh--Nagumo neuron is an essential condition for the occurrence of self-sustained oscillations in the neuron, i.e., one of the main modes of operation of the neuron. Effects related to dissipation in the FitzHugh--Nagumo neuron were considered in Refs.\cite{Freire2011,Yao2022}.

Time delays are a fundamental part of almost all biological phenomena. The finite propagation speed of the action potential along the axons of neurons and time lapses in information transmission (synaptic process) and reception (dendritic process) between neurons produce time delays in real neurons and their networks\cite{Tass2018}.      
The time required for neuronal communication can be significantly prolonged due to the physical distance between sending and receiving units\cite{Knoblauch2003,Knoblauch2004}, finite velocity of signal transmission\cite{Desmedt1980}, morphology of dendrites and axons\cite{Manor1991,Boudkkazi2007} and information processing time of the cell\cite{Wang2009}. 
Thus, time delays can play a crucial role in the dynamics of neuronal networks and should be taken into account in mathematical modeling and analysis\cite{stepan2009delay, Pet2019}. The role of delay in signal transmission in brain circuits is also worth to be noted\cite{Pariz2021}. Neglecting realistic time delays in mathematical models can lead to discrepancies between theoretical and experimental findings and thus prevents insight into relevant physiological mechanisms.

Many recent studies have been devoted to delay-induced phenomena\cite{Balanov2004, Balanov2006, Scholl2009, Popovych2011, Kantner2015} and effects of time delays on the synchronization dynamics of neuronal networks\cite{Choe2010,Kyr2011,Lehnert2011, Panchuk2013, Plotnikov2016, Wille2014, Esf2014, Kyr2014, Gjur2014,Esf2016, Pariz2018, Zia2020}. It is a well known, and often used, fact that time delay can destabilize a stationary point and introduce oscillatory behavior. The case
of two delay-coupled FitzHugh--Nagumo systems has been examined in detail showing that stable periodic oscillations may coexist with a stable steady state\cite{Scholl2009, Buric2003, Dahlem2009, Valles2011}. The bifurcation phenomenon is fully induced by the delay $\tau$ and represents a new form of oscillatory synchronization exhibiting a period close to $2\tau$. It has been reported that delay-enhanced synchronization may be essential for information transmission in neuronal networks\cite{Wang2009, Tang2011}. It has also been revealed that coupling delays present in the electrical or chemical synaptic connections can influence the synchronization of neuronal firing\cite{Dham2004, Yang2017}. In Ref.\cite{Wang2020} the synchronization phenomenon in a time-delayed chaotic system with unknown and uncertain parameters was studied and an intermittent adaptive control scheme was developed to guarantee synchronization between neurons. It has been investigated how the phase lag synchronization between the neurons of different brain regions is governed by the spatio-temporal organization of the brain by using self-sustained time-delayed chaotic oscillators\cite{Pet2019}.

%The dynamics of the complex system depends on the properties of each of the units and on their interactions.
With the discovery of special patterns of partial synchronization, called chimera states\cite{Kur2002, Abrams2004}, in networks of nonlocally coupled systems, a remarkable part of research was addressed to the effects of time delays on birth, stability  and control of chimera structures in various complex networks\cite{Larger2013, Semenov2016, Schoell2016, Ghosh2016, Gjur2017, Zakh2017, Saw2017, Saw2018, Saw2019, Nikitin2019, Saw2019a}. 
It has been shown that in networks with fractal connectivity, desired spatiotemporal
 patterns can be stabilized  by varying the time-delayed coupling between the nodes\cite{Saw2017, Saw2019a}. It has been demonstrated that delay allows to control amplitude chimeras\cite{Gjur2017} and coherence-resonance chimeras by adjusting delay time and feedback strength\cite{Zakh2017}. The interlayer delay in multilayer networks might act as a general organizing force to synchronize spatiotemporal patterns between layers. Recently, it has been shown that relay synchronization of chimeras in multiplex (triplex) networks can be controlled by time-delayed intra- and interlayer coupling\cite{Saw2018, Saw2019}.

In our work we use the paradigmatic example of the FitzHugh--Nagumo system in
the form, and for the parameter range, when the system displays excitable behavior. The form of the used equations differs from the simplified and usually considered FitzHugh--Nagumo system\cite{Scholl2009, Dahlem2009} since it takes into account a parameter which is responsible for the dissipation of a  neuron. The larger this parameter, the less the dissipation. In our numerical simulation of delay-coupled FitzHugh--Nagumo oscillators, we choose two different values of the dissipation parameter, one of which is close to the boundary of bistability and the other one is close to the borderline of the self-sustained oscillatory region. We first consider two delay-coupled neurons  and then extend our numerical simulation to a ring network with local and nonlocal coupling topology between the neurons. We study the interplay between dissipation in the single unit, the delay time, the coupling parameters and initial conditions  in forming the spatiotemporal dynamics of the networks. The numerical simulations performed and the results obtained can be straightforwardly generalized to the dynamics of delay-coupled simplified FitzHugh--Nagumo models\cite{Dahlem2009, Scholl2009, Zakharova2020}. 

\section{\label{sec:system}System under study}

As an object of our study we consider the FitzHugh--Nagumo oscillator which represents one of the simplest neuron models and is widely used in numerical simulation. This two-variable system is a paradigmatic model for neural excitability and is described as follows:
\begin{equation}
\begin{array}{l}
\varepsilon \dfrac{dx}{dt} = x - x^3/3 -y,\\[8pt]
\dfrac{dy}{dt} = \gamma x - y + \beta.
\end{array}
\label{eq:FHN_single}
\end{equation}
where $x$ is the fast variable (activator) and represents the voltage across the cell membrane, and $y$ is the slow recovery variable (inhibitor) which corresponds to the recovery state of the resting membrane of a neuron. All the control parameters are dimensionless. The small parameter $\varepsilon>0$ is the ratio of the activator to inhibitor time scales. The parameter $\beta$ determines the asymmetry and the parameter $\gamma$ is responsible for dissipation in the neuron. In an analog circuit of the FitzHugh--Nagumo oscillator (1) the parameter $\gamma$ is directly proportional to a ratio of two resistances ${R_0}/{R_6}$, where $R_6$ is the resistance of the oscillatory circuit\cite{Nguetcho2015}. The value of $R_6$ determines the amount of dissipation in the circuit. If the $R_6$ value increases, the dissipation increases and hence, the parameter $\gamma$ decreases, and vice versa. Both parameters $\beta$ and $\gamma$ also define the dynamical behavior of the neuron. Note that the form of FitzHugh--Nagumo equations~\eqref{eq:FHN_single} differs from the original form\cite{izhikevich2006fitzhugh, dahlem2008efficient, fitzhugh1961impulses}, but it is also investigated in a number of works\cite{nekorkin2008heteroclinic, kazantsev2001selective, nekorkin2005dynamics}. An analog electrical FitzHugh–Nagumo neuron, which is described by the system of equations like (\ref{eq:FHN_single}), was used in Ref.\cite{Nguetcho2015} to analyze its spiking responses on pulse stimulation. 

\begin{figure}[!t]
\centering
\includegraphics[width=0.995\linewidth]{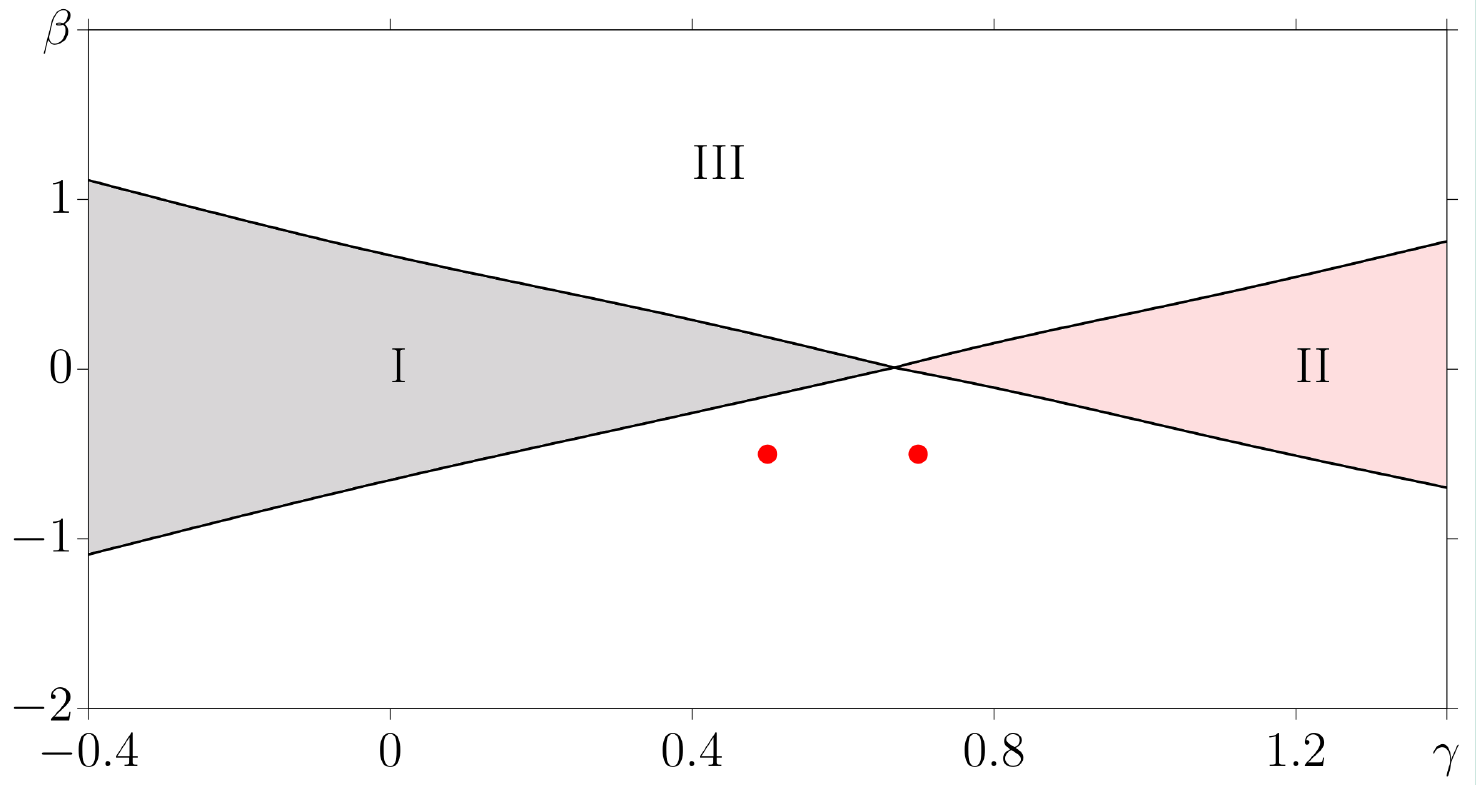}
\caption{Diagram of dynamical regimes of the single neuron~\eqref{eq:FHN_single} in the ($\gamma,\beta$) parameter plane for $\varepsilon = 0.01$. Regions I, II, and III correspond to the bistable (grey color), self-sustained oscillatory (pink color) and excitable regions (white color), respectively. In our paper we consider two sets of parameters of the single neuron marked by red dots which correspond to the excitable regime ($\gamma=0.5$ and $\gamma=0.7$ with $\beta=-0.5$).}
\label{fig:FHN_modes}
\end{figure}

The FitzHugh--Nagumo neuron model \eqref{eq:FHN_single} enables us to consider a variety of different dynamical behavior, namely excitable, self-sustained oscillatory, and bistable regimes. The ranges of existence of each dynamical regime are highlighted in the bifurcation diagram in the ($\gamma,\beta$) parameter plane shown in Fig.~\ref{fig:FHN_modes}.  The bistability region is labelled as I, the self-sustained oscillatory region is II, and the excitability region is denoted as III.
The solid lines in the diagram in Fig.~\ref{fig:FHN_modes} correspond to the fold bifurcation of the equilibrium points for $\gamma\lesssim 0.72$ and to the Hopf bifurcation for $\gamma\gtrsim 0.72$. The relevant values for fold bifurcations can be estimated analytically approximately as follows\cite{shepelev2017bifurcations}:
\[
\beta = \pm \dfrac{2}{3} \sqrt{1-\gamma}\left(1-\gamma\right).
\]
The location of the bifurcation lines in Fig.~\ref{fig:FHN_modes} is independent of the parameter $\varepsilon$. Inside region III (Fig.~\ref{fig:FHN_modes}), the neuron settles to the excitable regime when there are no self-sustained oscillations without an external force or perturbation. A single stimulus of a certain intensity can excite the neuron activity but it quickly decays and the system returns to the equilibrium\cite{shepelev2017bifurcations}.

%%%%%%%%%%%%%%%%%%%%%%%%%%%%%%% fig.1 %%%%%%%%%%%%%%%%%%%%%%%%%%%%%%%%%
%%%%%%%%%%%%%%%%%%%%%%%%%%%%%%%%%%%%%%%%%%%%%%%%%%%%%%%%%%%%%%%%%%%%%%%%%

%%%%%%%%%%%%%%%%%%%%%%%%%%%%%%% fig.2 %%%%%%%%%%%%%%%%%%%%%%%%%%%%%%%%%
\begin{figure}[!t]
\centering
\includegraphics[width=0.495\linewidth]{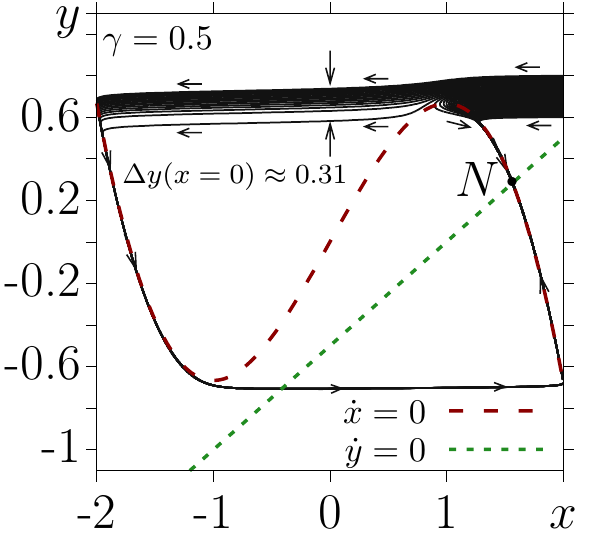}
\includegraphics[width=0.495\linewidth]{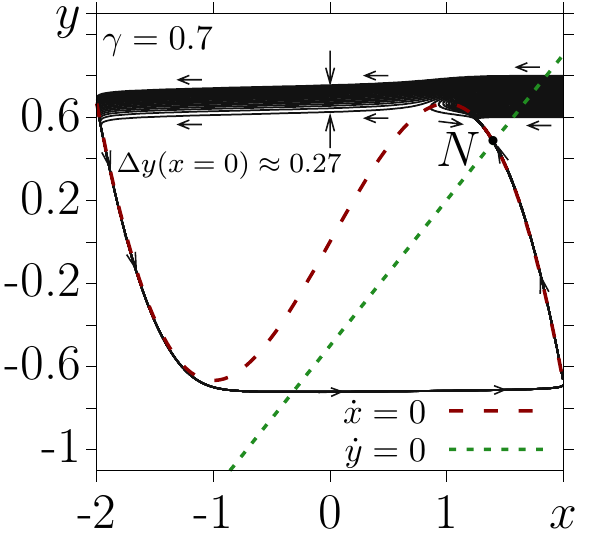}
\\
(a)\hspace{113pt}(b)
\\
\includegraphics[width=0.495\linewidth]{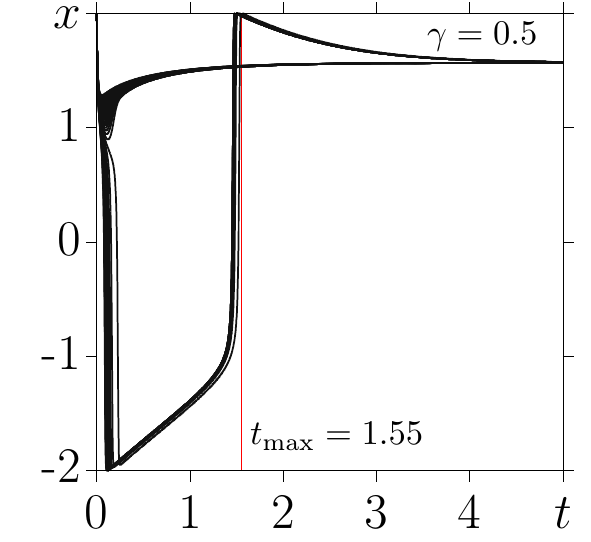}
\includegraphics[width=0.495\linewidth]{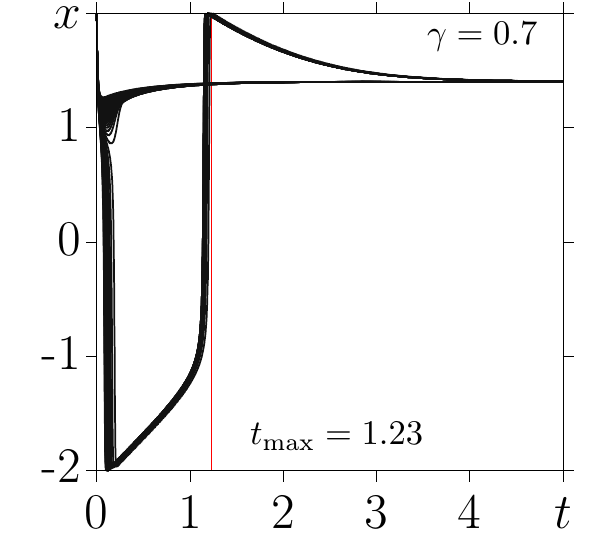}
\\
(c)\hspace{113pt}(d)
\caption{Phase portraits (a,b) and time series (c,d) for the system~\eqref{eq:FHN_single} at $\gamma=0.5$ (a,c) and $\gamma=0.7$ (b,d) with $\varepsilon=0.01$ and $\beta=-0.5$. The $x$ and $y$ null-isoclines are depicted as red dashed and green dotted curves, respectively. Several phase trajectories approaching the stable node $N$ for different initial conditions are plotted in black. The arrows show the direction of motion. The vertical red lines in the time series plots (c,d) show the time when the maximum spike amplitude is achieved.}
\label{fig:FHN_modesPP}
\end{figure}
%%%%%%%%%%%%%%%%%%%%%%%%%%%%%%%%%%%%%%%%%%%%%%%%%%%%%%%%%%%%%%%%%%%%%%%%%

In the following, in order to study how the parameter of dissipation can effect the dynamics of delay-coupled FitzHugh--Nagumo neurons we choose two different values of $\gamma$ marked by red dots in Fig.~\ref{fig:FHN_modes}. One of them, $\gamma=0.5$, is close to the boundary of bistability (Fig.~\ref{fig:FHN_modes}) and the other one, $\gamma=0.7$, is closer to the boundary of the self-sustained oscillatory regime. We recall that the second case corresponds to smaller dissipation in a single neuron. Throughout the paper, we fix the values of $\varepsilon=0.01$ and $\beta=-0.5$. This corresponds to the excitable regime observed in a single FitzHugh--Nagumo oscillator within a rather wide range of $\gamma \in[0.2,1.1]$ (see Ref.\cite{shepelev2017bifurcations}).
Figure~\ref{fig:FHN_modesPP}(a,b) shows phase trajectories approaching the stable node $N$ from a set of different initial conditions for the system~\eqref{eq:FHN_single} for fixed $\beta$ and $\varepsilon$ and for the two chosen values of $\gamma$. As can be seen, in the case of smaller dissipation (Fig.~\ref{fig:FHN_modesPP}(b)), the phase trajectories starting with different initial states lie more closely and densely to each other as compared with the case of larger dissipation (Fig.~\ref{fig:FHN_modesPP}(a)). Some of the trajectories shown in Fig.~\ref{fig:FHN_modesPP} move along the big excitability loop in the directions indicated by the arrows. As can be seen from the time series plotted in Fig.~\ref{fig:FHN_modesPP}(c,d), the time for achieving a maximum spike amplitude is smaller when dissipation is low ($t_{\rm max}=1.23$ for $\gamma=0.7$). As a consequence, the trajectories converge  towards the stable node $N$ more slowly. In the case of a higher level of dissipation  (Fig.~\ref{fig:FHN_modesPP}(c)), a larger time is taken to show the maximum spike amplitude ($t_{\rm max}=1.55$ for $\gamma=0.5$), and the trajectories approach the stable node faster than compared with a smaller dissipation.

% As the parameter $\gamma$ in the FHN system is concerned, then it is easy to see that its increase leads to a decrease of the resistance ratio in the system and thus to a decrease of dissipation. In an dissipative oscillatory circuit, on the contrary, increasing $\gamma$ causes dissipation to increase. Besides, an isolated autonomous FHN neuron for both values of $\gamma$ (0.5 and 0.7) exhibits a single stable equalibrium - a stable focus. However, the trajectory starting with the same initial condition converges to the focus faster at $\gamma=0.5$ than for $\gamma=0.7$. In the context of the dissipative oscillatory circuit, such a behavior occurs when the resistance decreases. 

%Если рассмотреть параметр гамма, то несложно увидеть, что его увеличение ведет к уменьшению отношения сопротивлений в системе, и, следовательно к уменьшению диссипации. Увеличение значения гамма, наоборот, ведет к увеличению диссипации в колебательном контуре. Кроме того, если рассмотреть изолированный и автономный нейрон ФитцХью-Нагума при значениях гамма = 0.8 и 0.5, то можно увидеть, что единственное устойчивое состояние равновесия при этих параметрах представляет собой устойчивый фокус. Однако, с одного и того же начального условия при гамма = 0.5 траектория сходится к точке равновесия значительно быстрее, чем при гамма = 0.8. Если проводить аналогию с диссипативным колебательным контуром, то в нем такое поведение наблюдается при уменьшении сопротивления.

\section{Two delay-coupled FitzHugh--Nagumo neurons}

Since time delay is inevitable due to the finite propagating speed in the signal transmission between the neurons, we start with considering two identical FitzHugh--Nagumo neurons with time-delayed coupling and explore how the self-sustained oscillatory activity of this system depends on the delay time, the coupling strength, and the parameter of dissipation of the single neuron.

For bidirectional delayed coupling between the two FitzHugh--Nagumo oscillators, we have the following system of equations:
\begin{equation}
\begin{array}{l}
\varepsilon \dfrac{dx_1}{dt} = x_1 - x_1^3/3 -y_1 + \sigma (x_2(t-\tau)-x_1),\\[8pt]
\dfrac{dy_1}{dt} = \gamma x_1 - y_1 + \beta,\\[8pt]
\varepsilon \dfrac{dx_2}{dt} = x_2 - x_2^3/3 -y_2 + \sigma (x_1(t-\tau)-x_2),\\[8pt]
\dfrac{dy_2}{dt} = \gamma x_2 - y_2 + \beta,
\end{array}
\label{eq:twoFHN}
\end{equation}
where $x_i$ and $y_i$, $i=1,2$ are the dynamical variables defined in Eq.~\eqref{eq:FHN_single} for a single neuron. The coupling strength between the oscillators is determined by the parameter $\sigma$, and  $\tau>0$ represents the time delay in signal transmission. This system with the simplified FitzHugh--Nagumo oscillator form $\dot{y}_i=x_i+\beta$ has already been studied in\cite{Scholl2009, Dahlem2009, Nikitin2019}.

\subsection{Delay-induced oscillations}

%%%%%%%%%%%%%%%%%%%%%%%%%%%%%%% fig.3 %%%%%%%%%%%%%%%%%%%%%%%%%%%%%%%%%
\begin{figure}[!t]
\centering
\includegraphics[width=0.495\linewidth]{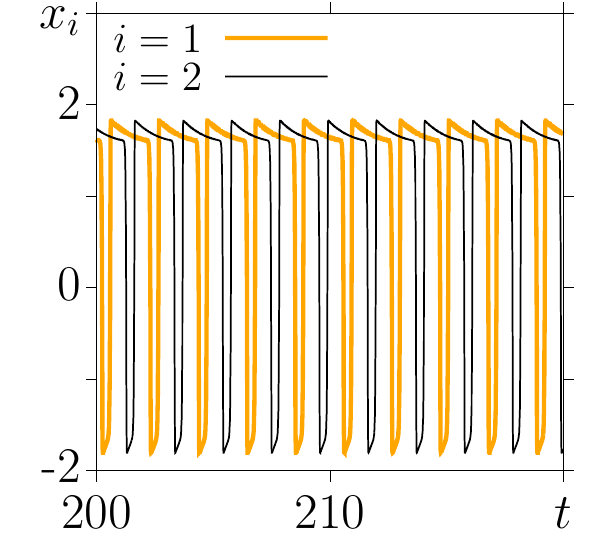}
\includegraphics[width=0.495\linewidth]{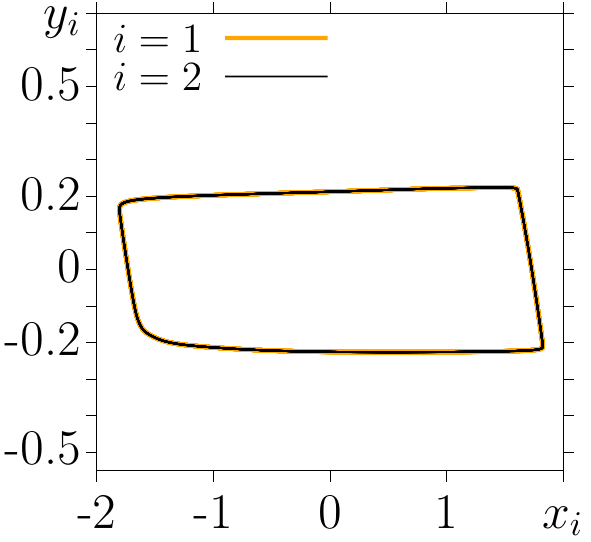}
(a1) \hspace{25pt} $\gamma=0.5,\tau=1$ \hspace{25pt} (a2)
\\
\includegraphics[width=0.495\linewidth]{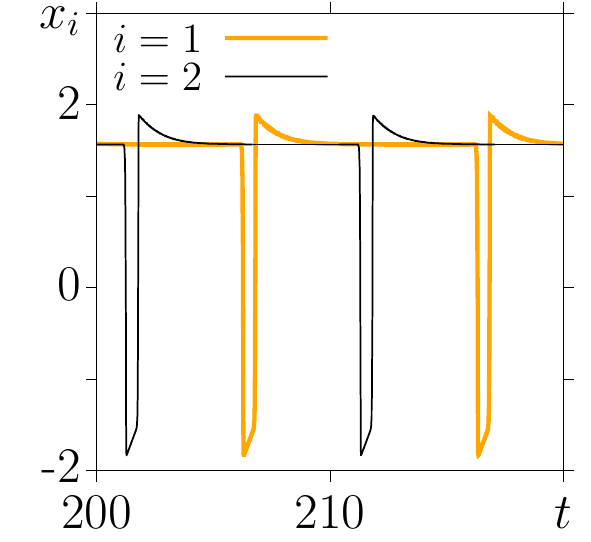}
\includegraphics[width=0.495\linewidth]{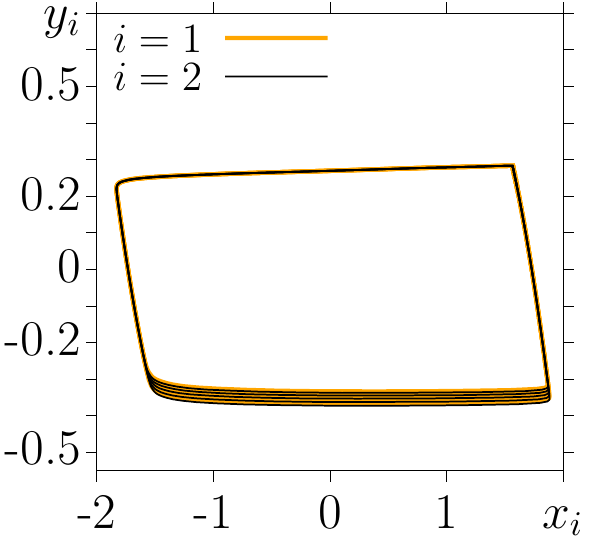}
(b1) \hspace{25pt} $\gamma=0.5,\tau=5$ \hspace{25pt} (b2)
\\
\includegraphics[width=0.495\linewidth]{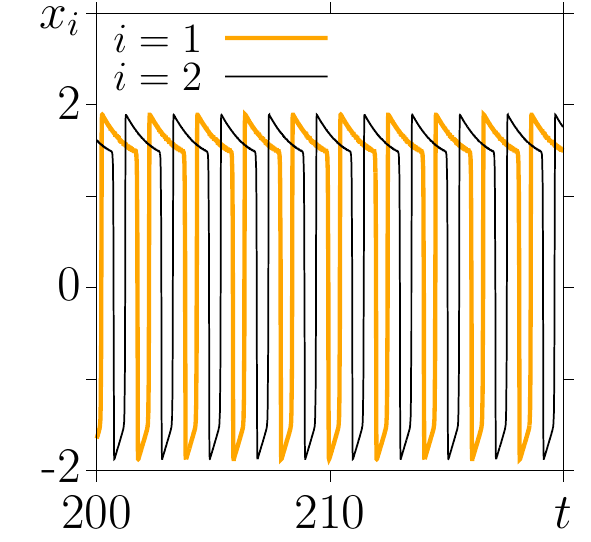}
\includegraphics[width=0.495\linewidth]{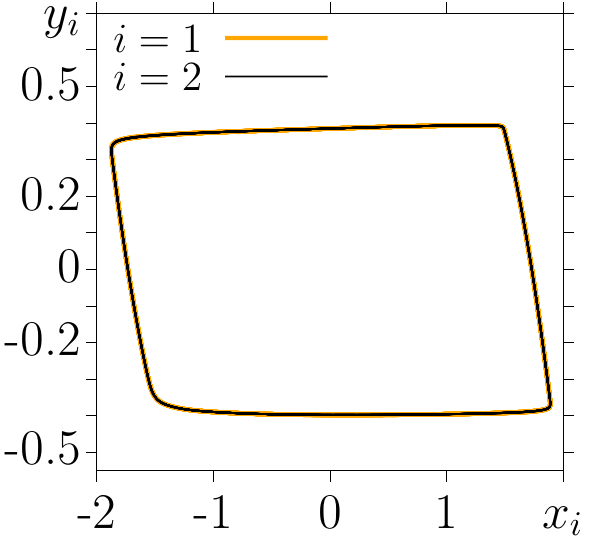}
(c1) \hspace{25pt} $\gamma=0.7,\tau=1$ \hspace{25pt} (c2)
\\
\includegraphics[width=0.495\linewidth]{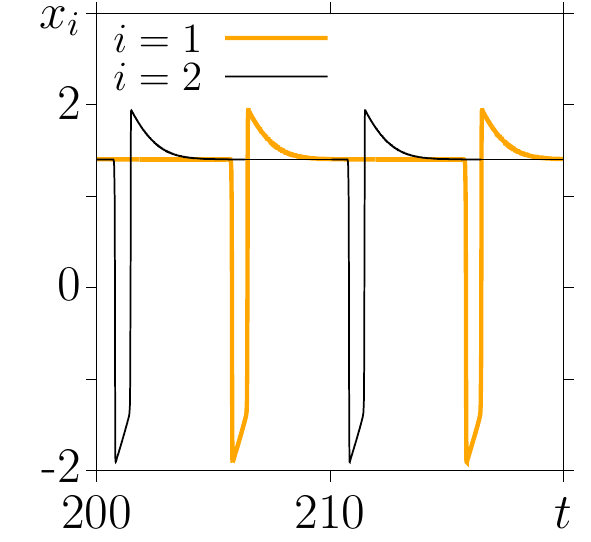}
\includegraphics[width=0.495\linewidth]{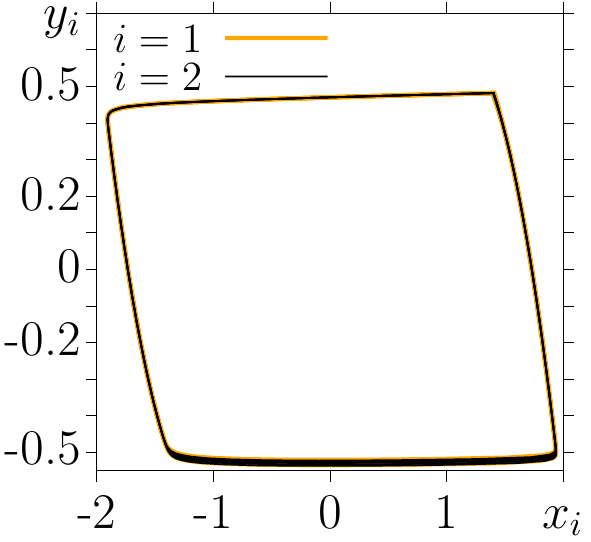}
(d1) \hspace{25pt} $\gamma=0.7,\tau=5$ \hspace{25pt} (d2)
\caption{Anti-phase oscillations for the two delay-coupled FitzHugh--Nagumo neurons~\eqref{eq:twoFHN} for $\gamma=0.5$ (panels (a1), (a2), (b1), (b2)) and  $\gamma=0.7$ (panels (c1), (c2), (d1), (d2)) and for the time delay $\tau=1$ (panels (a1), (a2), (c1), (c2)) and $\tau=5$ (panels (b1), (b2), (d1), (d2)). Time series for the 1st ($i=1$, black curves) and 2nd ($i=2$, orange curves) neurons are shown in panels (a1), (b1), (c1) and (d1). The corresponding phase portrait projections are illustrated in panels (a2), (b2), (c2) and (d2).  Initial conditions correspond to the case when only one neuron is excited. Other parameters: $\sigma=0.3$, $\varepsilon=0.01$, and $\beta=-0.5$.}
\label{fig:2neuron_osc}
\end{figure}
%%%%%%%%%%%%%%%%%%%%%%%%%%%%%%%%%%%%%%%%%%%%%%%%%%%%%%%%%%%%%%%%%%%%%%%%%

%Within this work, we fix  $\varepsilon=0.01$ and $\beta=-0.5$, which correspond to the excitable regime observed in a single FitzHugh--Nagumo oscillator within a rather wide range of $\gamma \in[0.2,1.1]$ (see Ref.~\cite{shepelev2017bifurcations}).
The dynamics of the system~\eqref{eq:twoFHN} is explored for the parameter values indicated in Sec.~I. 
%Both cases are related to a sufficiently strong dissipation and to the regimes located away from the boundary of the Hopf bifurcation.
For numerical simulation within this work we use the Runge-Kutta method with time step $h=0.005$. The maximum Lyapunov exponent is defined here as
\begin{equation}
\Lambda(t_{0},T):=\dfrac{1}{T}\ln\dfrac{\|\boldsymbol{\xi}(t_{0}+T)\|}{\|\boldsymbol{\xi}(t_{0})\|},\label{eq:LE-1}
\end{equation}
and represents the finite-time growth rate of a small perturbation $\boldsymbol{\xi}(t_{0})$ introduced in the variational differential equations. The initial conditions are given in the form of two random values in both neurons. The initial states (history function) of the coordinates are chosen as follows: $x_{1,2}(t\in[-\tau;0])=x_{1,2}^0$ and $y_{1,2}(t\in[-\tau;0])=y_{1,2}^0$, where $x_{1,2}^0 \in [1,2]$ and $y_{1,2}^0 \in [0,1]$ to induce the in-phase oscillatory regime, and from the ranges $x_1^0 \in [1,2], y_1^0 \in [0,1], x_2^0 \in [-2,-1], y_2^0 \in [-1,0]$ to obtain the anti-phase oscillatory regime. Without delay and for any values of the coupling strength $\sigma$ the two neurons demonstrate oscillations during only a very short transient time, after which they switch to the same equilibrium point without any oscillations. When the delayed coupling is introduced between the neurons, periodic self-sustained oscillations are excited in the system starting with a certain threshold of the coupling strength ($\sigma\approx 0.2$ for $\gamma=0.5,\tau\gtrsim 1$ and $\sigma\approx 0.1$ for $\gamma=0.7,\tau\gtrsim 1$). Depending on initial conditions the neurons demonstrate both complete in-phase synchronization and complete anti-phase synchronization with each other.

The neuronal dynamics is illustrated in Fig.~\ref{fig:2neuron_osc} for two different values of the dissipation parameter ($\gamma=0.5$ and $\gamma=0.7)$, the time delay ($\tau=1$ and $\tau=5$), and for the anti-phase states. In this case the initial conditions are chosen in such a way that only one neuron is excited ($x_1^0 \in [1,2], y_1^0 \in [0,1], x_2^0 \in [-2,-1], y_2^0 \in [-1,0]$). Panels (a1)-(b2) in Fig.~\ref{fig:2neuron_osc} show results for $\gamma=0.5$, and (c1)-(d2) illustrate regimes for $\gamma=0.7$. Time series for variables $x_{i}(t)$, $i=1,2$ (black and orange curves, respectively) are plotted in panels with index 1, and the corresponding projections of phase portraits for both neurons are pictured in panels with index 2. It is seen that despite the excitable regimes of the isolated neurons and the absence of any external forces, the delayed coupling leads to the emergence of periodic self-sustained oscillations (the maximal Lyapunov exponent $\Lambda$ is equal to 0). This is a well-known result and has been reviewed in Ref.~\cite{Scholl2009}. The temporal behavior for both neurons reflects anti-phase spike oscillations (Fig.~\ref{fig:2neuron_osc}(a1)-(d1)). They are characterized by a slow motion around the equilibrium (in the isolated neuron) and a fast motion along a limit cycle that is illustrated in Fig.~\ref{fig:2neuron_osc}(a2)-(d2) for different values of the parameters. Similar results for two delay-coupled FitzHugh--Nagumo neurons in the simplified form have been obtained numerically\cite{Scholl2009} and analytically\cite{Nikitin2019}. Besides, the oscillations of the first and second neurons correspond to the same limit cycle. For a larger value of  $\gamma=0.7$ (Fig.~\ref{fig:2neuron_osc}(c1,c2,d1,d2)), the size of the limit cycle extends in the $y$-axis, while it remains almost unchanged along the $x$-axis. It is related to the slope of the $y$-nullcline as $\gamma$ grows. Hence, the $x$- and $y$-nullclines intersect for larger values of the $y$ variable, and the oscillation amplitude in this variable becomes larger (compare Figs.~\ref{fig:2neuron_osc} and~\ref{fig:FHN_modesPP}).

We now analyze how the neuronal dynamics changes if we vary the time delay $\tau$ in the system \eqref{eq:twoFHN}.
 Comparison of the time series plotted in Fig.~\ref{fig:2neuron_osc}(a1,c1) for $\tau=1$ and in Fig.~\ref{fig:2neuron_osc}(b1,d1) for $\tau=5$ gives evidence that the oscillation period increases when the time delay becomes larger. As can be seen from the plots, this result is independent of the parameter of dissipation. 
%{\color{red} Besides, weak quasi-periodic oscillations can be induced for a larger value of $\tau$, as follows from the phase portrait projections in Fig.~\ref{fig:2neuron_osc},(b2,d2). \color{blue} E.S.: I do not think this is quasi-periodic, rather it may be due to numerical inaccuracy. If we claim this is quasi-periodic, we must present evidence, e.g., power spectra with two incommensurate frequencies. \color{red} Regular oscillations!}
Our calculations show that the oscillation period $T$ strongly depends on the delay $\tau$ as shown analytically in the simplified FitzHugh--Nagumo model\cite{Scholl2009, Nikitin2019}. Moreover, we reveal that the period is always equal to 2$\tau$ for the chosen initial conditions. This effect is rather easy to explain\cite{Dahlem2009, Panchuk2013}. Since the initial conditions are chosen randomly, only a single neuron is excited, while the other one is quiescent after a short time. After the time delay, the second (quiescent) neuron gets a signal from the first (excited) neuron and is also excited. In turn, the first neuron switches to the quiescent regime until it receives a signal from the second neuron via the coupling. Hence, the oscillations in the system are anti-phase to each other and the period is strictly equal to the doubled time delay. It should be noted that the emergence of delay-induced oscillations has a threshold character. This means that the oscillations occur only when the coupling strength $\sigma$ exceeds a certain threshold. It is reasoned by the following fact. The intensity of a signal which a neuron receives from its neighbor is equal to $\sigma A$, where $A$ is the oscillation amplitude. This signal can be considered as an external  periodic force. In this case, the neuron demonstrates its activity only if the external force amplitude is larger than a certain threshold level. It is worth noting that the dynamics shown in Fig.~\ref{fig:2neuron_osc} is reminiscent of the reverberation of the activity between the two nodes. An interesting application could be the relation with the propagation of the signals in layered neuronal networks explored in Ref.~\cite{Rezaei2020}.
 
%%%%%%%%%%%%%%%%%%%%%%%%%%%%%%% fig.4 %%%%%%%%%%%%%%%%%%%%%%%%%%%%%%%%%
\begin{figure}[!t]
\centering
\includegraphics[width=0.495\linewidth]{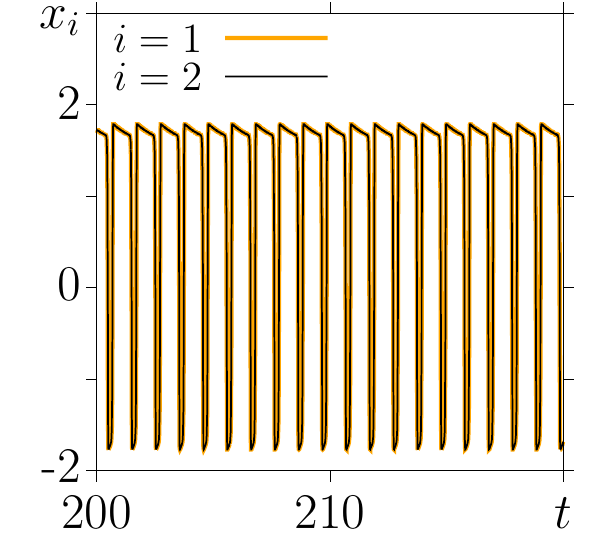}
\includegraphics[width=0.495\linewidth]{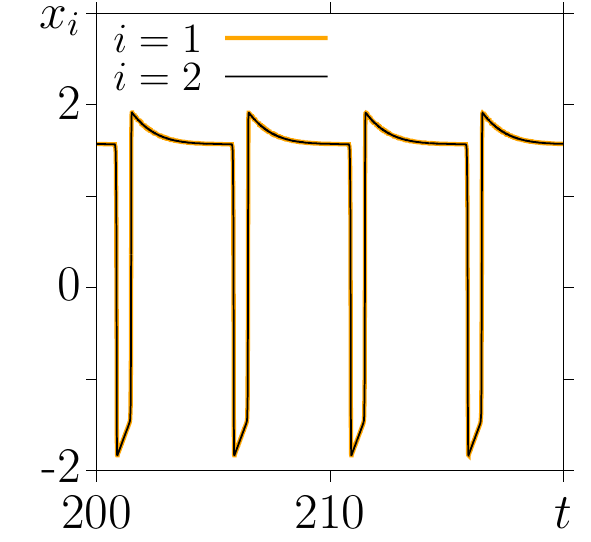}
\\
(a)\hspace{113pt}(b)
\\
\includegraphics[width=0.495\linewidth]{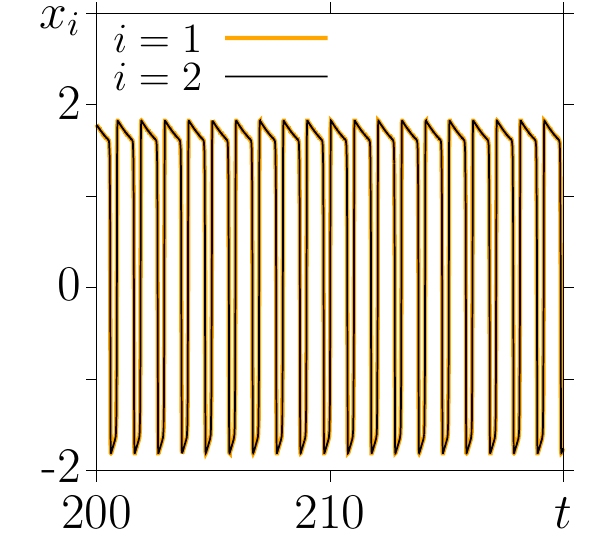}
\includegraphics[width=0.495\linewidth]{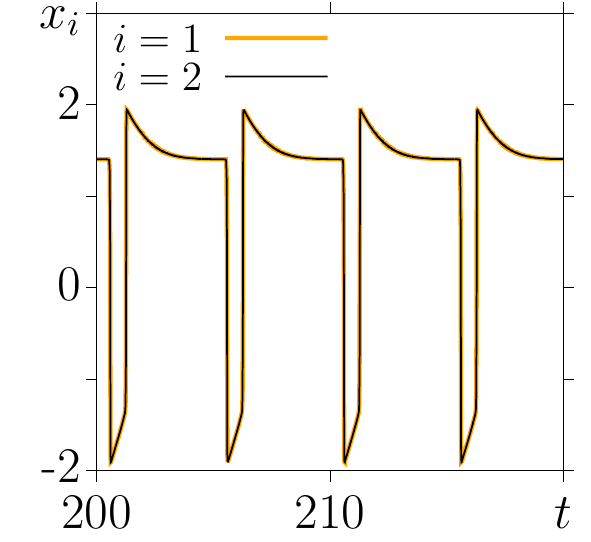}
\\
(c)\hspace{113pt}(d)
\caption{In-phase oscillations for the two delay-coupled FitzHugh--Nagumo neurons~\eqref{eq:twoFHN} for $\gamma=0.5$ (a,b) and $\gamma=0.7$ (c,d). Time series for the 1st ($i=1$, black curves) and 2nd  ($i=2$, orange curves) neurons are plotted for $\tau=1$ (a,c) and $\tau=5$ (b,d).  Initial conditions correspond to the case when both neurons are initially excited. Other parameters: $\varepsilon=0.01$, $\sigma=0.5$, and $\beta=-0.5$.}
\label{fig:2neuron_osc_inPhase}
\end{figure}
%%%%%%%%%%%%%%%%%%%%%%%%%%%%%%%%%%%%%%%%%%%%%%%%%%%%%%%%%%%%%%%%%%%%%%%%%
 
The next question is what changes will happen if both neurons are initially excited? In this case, the neurons begin to oscillate in-phase starting with a certain value of $\sigma$. Examples of the time series $x_{i}(t)$, $i=1,2$, are depicted in Fig.~\ref{fig:2neuron_osc_inPhase} for two values of dissipation $\gamma=0.5$ (Fig.~\ref{fig:2neuron_osc_inPhase}(a,b)) and $\gamma=0.7$ (Fig.~\ref{fig:2neuron_osc_inPhase}(c,d)) and for the time delay $\tau=1$ (Fig.~\ref{fig:2neuron_osc_inPhase}(a,c)) and $\tau=5$ (Fig.~\ref{fig:2neuron_osc_inPhase}(b,d)). It is seen that now both neurons oscillate synchronously in-phase. In the case of smaller dissipation (Fig.~\ref{fig:2neuron_osc_inPhase}(c,d)), the form of spikes in the time series becomes more pronounced and sharper for both time delay values as compared with the corresponding spike sequences for a larger dissipation (Fig.~\ref{fig:2neuron_osc_inPhase}(a,b)). 
Comparing the time series obtained for the same time delay in Fig.~\ref{fig:2neuron_osc}(a1,c1) (anti-phase oscillations) and  Fig.~\ref{fig:2neuron_osc_inPhase}(a,b) (in-phase oscillations) indicates that the frequency is doubled for the anti-phase oscillations. At the same time, the projections of phase portraits are completely equal to the previous cases depicted in Fig.~\ref{fig:2neuron_osc}(a2,b2). The observed differences occur due to a different mechanism of delay-induced oscillations. For the first choice of initial conditions, the neurons are excited consistently, one after another, with the oscillation period $T=2\tau$. For the second case of initial conditions, both neurons are activated simultaneously with the period being equal to the delay. For the general discussion of such exchange of pulses, including delayed feedbacks with different delay times see Ref.~\cite{Panchuk2013}. 

%%%%%%%%%%%%%%%%%%%%%%%%%%%%%%% fig.5 %%%%%%%%%%%%%%%%%%%%%%%%%%%%%%%%%
\begin{figure}[!t]
\centering
\includegraphics[width=0.495\linewidth]{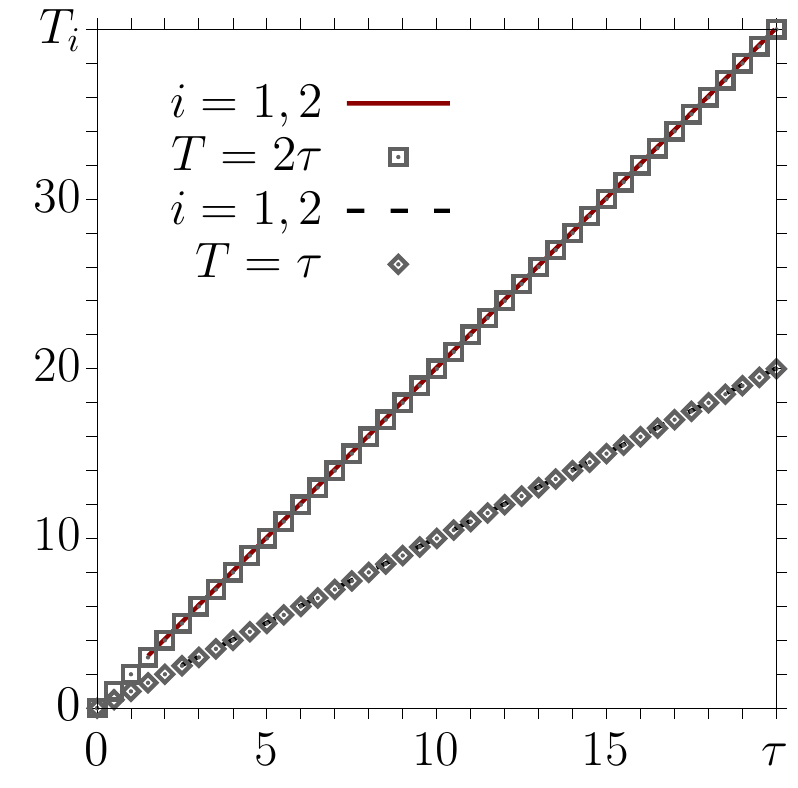}
\includegraphics[width=0.495\linewidth]{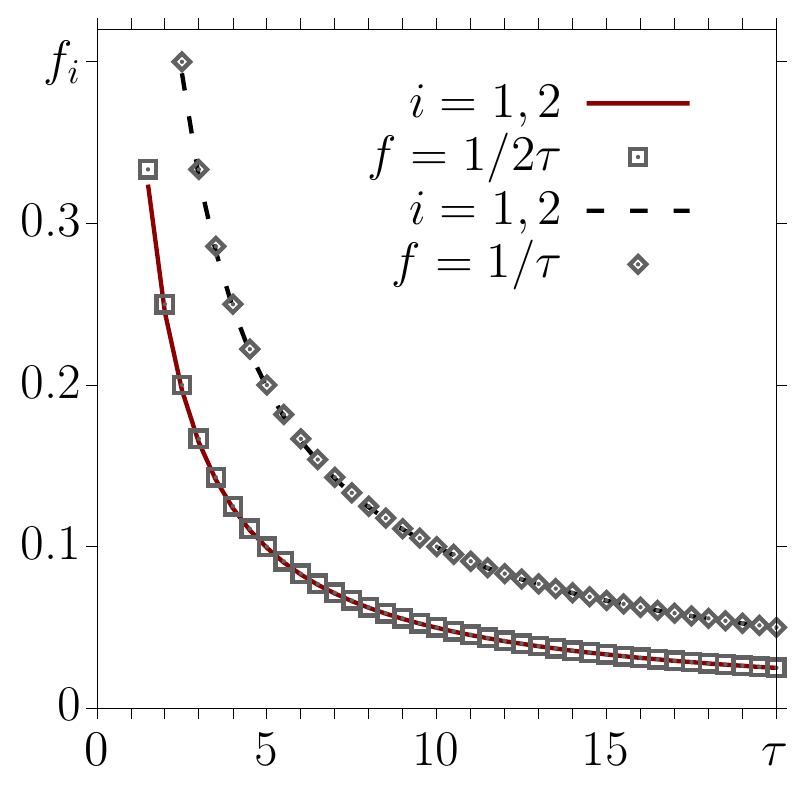}
\\
(a)\hspace{113pt}(b)
\caption{Dependences of the period $T_i$ (a) and frequency $f_i$ (b) for two neurons ($i=1,2$) on the time delay $\tau$ for anti-phase oscillations (red curves) and  in-phase oscillations (black curves) in the system \eqref{eq:twoFHN} for $\gamma=0.5$. Functions $T(\tau)=\tau$ and $f(\tau)=1/\tau$ are shown by squares and  $T(\tau)=2\tau$ and $f(\tau)=1/2\tau$ by circles. Other parameters: $\varepsilon=0.01$, $\sigma=0.5$, and $\beta=-0.5$.}
\label{fig:2neuron_freq}
\end{figure}
%%%%%%%%%%%%%%%%%%%%%%%%%%%%%%%%%%%%%%%%%%%%%%%%%%%%%%%%%%%%%%%%%%%%%%%%%

\begin{figure}[!t]
\centering
\includegraphics[width=0.95\linewidth]{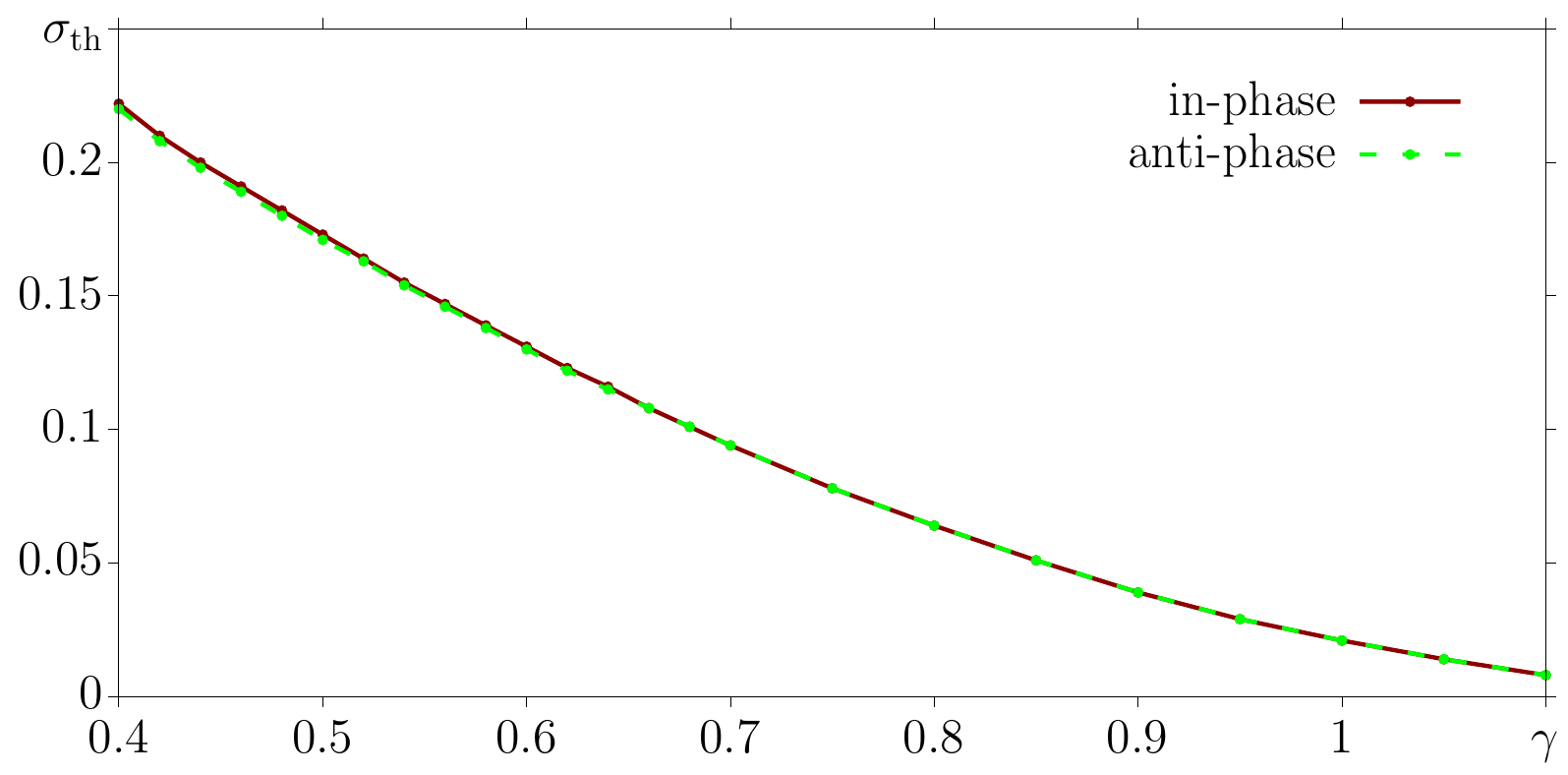}
\caption{Dependence of the threshold for the oscillation excitation $\sigma_{\text{th}}$ versus the dissipation parameter $\gamma$ in the system of two delay-coupled neurons for in-phase (red curve with dots) and anti-phase (green dashed curve with dots) oscillations at $\tau=5$. Other parameters: $\varepsilon=0.01$, and $\beta=-0.5$.}
\label{fig:2neuron_threshold}
\end{figure}

We now evaluate how the period and the frequency of oscillations depend on the delay for both types of initial conditions mentioned above. The calculated dependences are plotted in Fig.~\ref{fig:2neuron_freq} for $\gamma=0.5$ and for anti-phase (red curves) and in-phase (black curves) oscillations. As follows from the graphs in Fig.~\ref{fig:2neuron_freq}(a), the period $T_i$ of oscillations in both neurons labeled with $i=1,2$ is strictly equal to the doubled time delay $2\tau$ for the anti-phase oscillations and to $\tau$ for the in-phase oscillatory regime. In both cases the period $T_i$ increases monotonically and linearly as the delay time $\tau$ grows. In turn, the frequency $f_i$ presents the function inverse of the time delay $1/(2\tau)$ for the anti-phase oscillations and $1/\tau$ for the in-phase oscillations (Fig.~\ref{fig:2neuron_freq}(b)). This is in line with the results in\cite{Scholl2009, Dahlem2009, Nikitin2019, Panchuk2013}.  Moreover, our calculations show that both the period and the frequency of oscillations are defined only by the delay and are independent of the coupling strength $\sigma$ and the parameter $\gamma$. However, there is a certain threshold with respect to $\sigma$, when the oscillations are excited, and this value depends on  the delay and the control parameters, see analytical results in Refs.\cite{Nikitin2019, Panchuk2013}. Our calculations show that the dissipation parameter $\gamma$ can influence the threshold for the neural activity in the system \eqref{eq:twoFHN}. Dependences of the $\sigma$ threshold value on $\gamma$ are plotted in Fig.~\ref{fig:2neuron_threshold}) for fixed delay time $\tau=5$ and for the in-phase and anti-phase oscillations. The parameter $\gamma$ is varied within the region of excitable dynamics (region III in Fig.~\ref{fig:FHN_modes}). It is seen that both dependences coincide and the threshold value gradually decreases as the parameter $\gamma$ increases, i.e., as the dissipation level in the neuron decreases.

%%%%%%%%%%%%%%%%%%%%%%%%%%%%%%% fig.5c %%%%%%%%%%%%%%%%%%%%%%%%%%%%%%%%%
%%%%%%%%%%%%%%%%%%%%%%%%%%%%%%%%%%%%%%%%%%%%%%%%%%%%%%%%%%%%%%%%%%%%%%%%%

\begin{figure}[!t]
\centering
$\gamma=0.5$ \hspace{93pt} $\gamma=0.7$
\\
\includegraphics[width=0.495\linewidth]{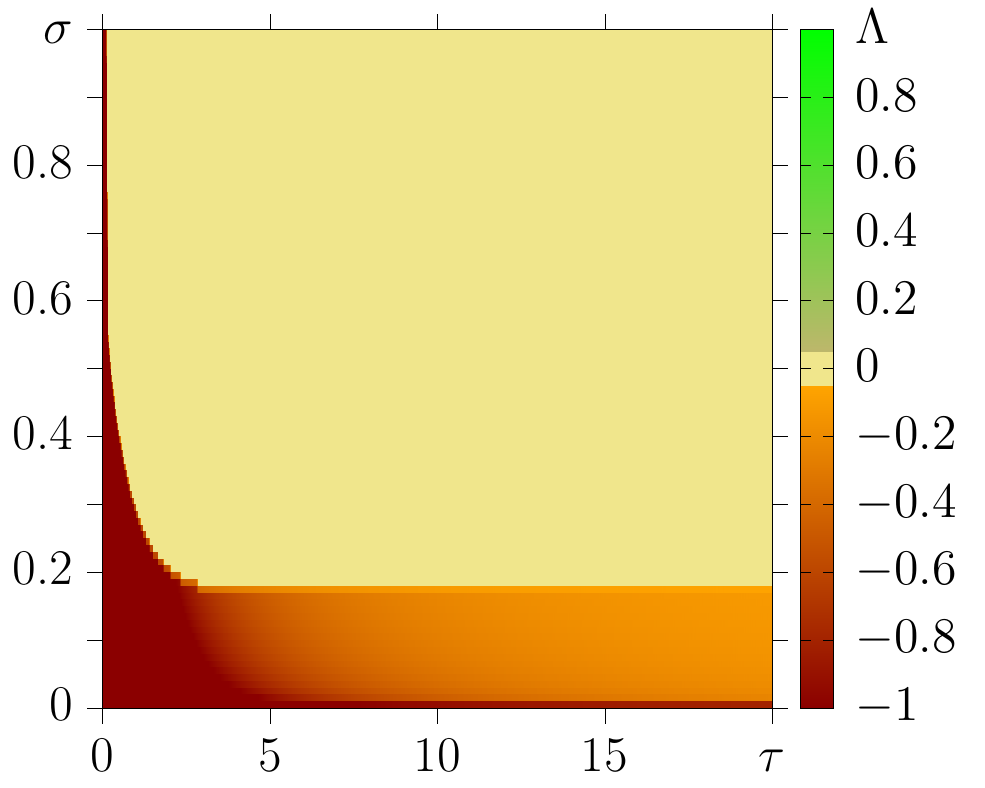}
\includegraphics[width=0.495\linewidth]{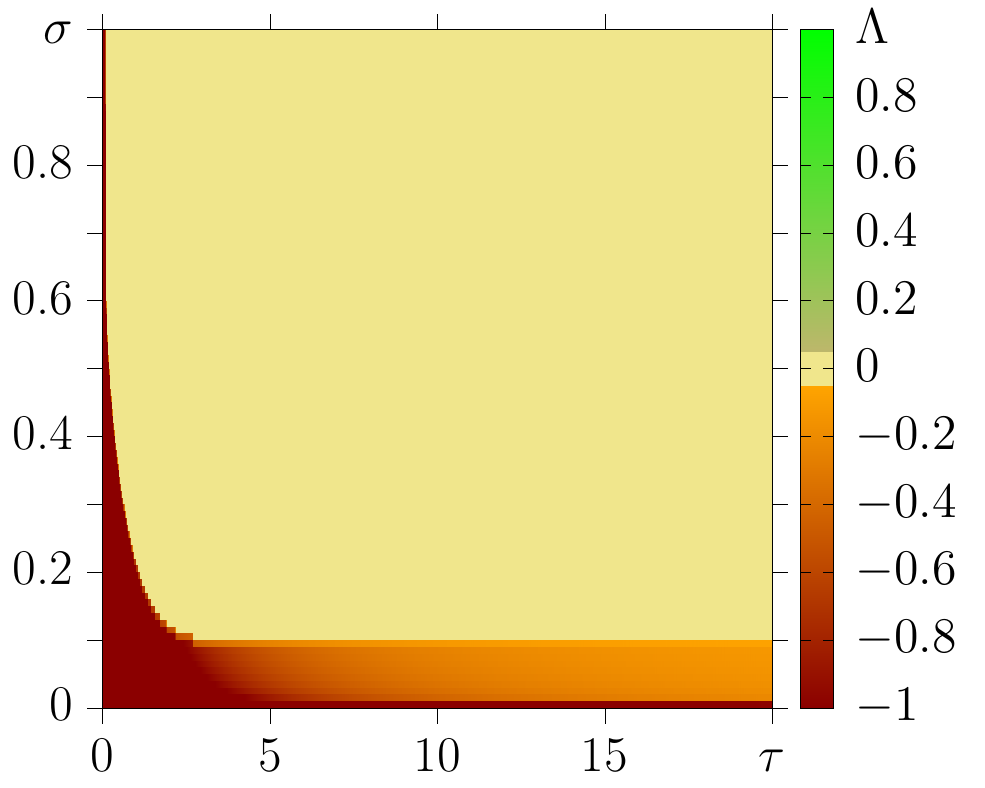}
\\
(a)\hspace{13pt} anti-phase oscillations \hspace{13pt}(b)
\\
\includegraphics[width=0.495\linewidth]{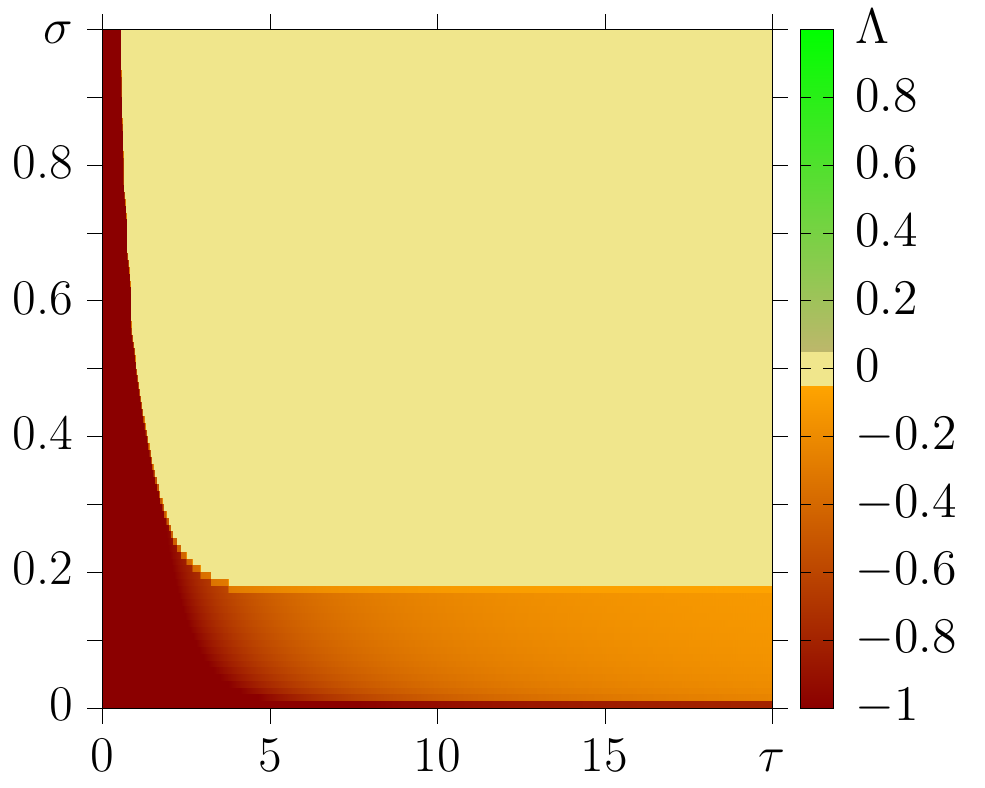}
\includegraphics[width=0.495\linewidth]{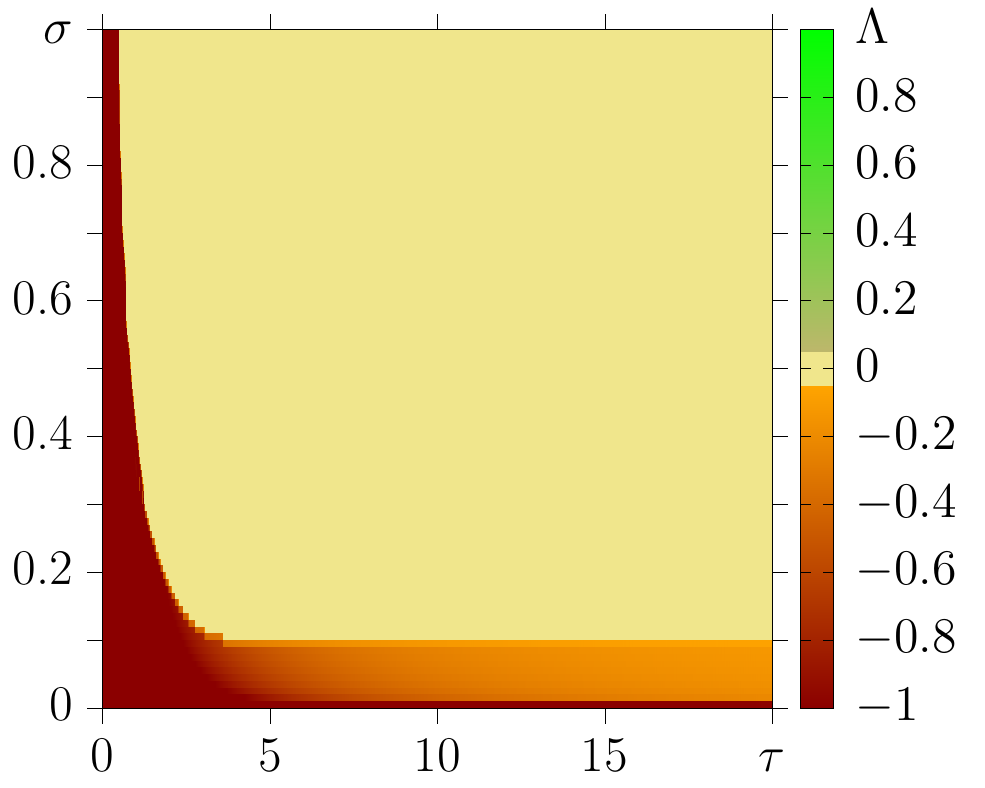}
\\
(c)\hspace{17pt} in-phase oscillations \hspace{17pt}(d)
\caption{Distributions of the maximal Lyapunov exponent $\Lambda$ in the ($\tau,\sigma$)  parameter plane for anti-phase oscillations (upper row) and in-phase oscillations (lower row) at $\gamma=0.5$ (a,c) and $\gamma=0.7$ (b,d). Other parameters: $\varepsilon=0.01$, $\beta=-0.5$.}
\label{fig:2neuron_LE}
\end{figure}

To evaluate the region of existence of delay-induced oscillations in the system~\eqref{eq:twoFHN}, we calculate the maximal Lyapunov exponent $\Lambda$ and plot its values in the ($\tau,\sigma$) parameter plane for two different values of $\gamma=0.5$ and $\gamma=0.7$. The corresponding diagrams are presented in Fig.~\ref{fig:2neuron_LE}(a,b) for anti-phase oscillations and in Fig.~\ref{fig:2neuron_LE}(c,d) for in-phase oscillations.

%%%%%%%%%%%%%%%%%%%%%%%%%%%%%%% fig.6 %%%%%%%%%%%%%%%%%%%%%%%%%%%%%%%%%
%%%%%%%%%%%%%%%%%%%%%%%%%%%%%%%%%%%%%%%%%%%%%%%%%%%%%%%%%%%%%%%%%%%%%%%%%

Obviously, the values of $\Lambda$ must be negative when there are no oscillations and zero or positive (only for chaos) for the oscillatory regime. In our case the delay-induced oscillations are always non-chaotic. Hence, the maximal Lyapunov exponent is always zero. Thus, the distribution of $\Lambda$ in the $(\tau,~\sigma)$ parameter plane enables one to distinguish the quiescent (red color, negative values of $\Lambda$) and the oscillatory (yellow color, $\Lambda=0$) regimes and thus to define the boundary between them. The diagrams show that there is a threshold for the oscillation excitation with respect to both the coupling strength $\sigma$ and the delay $\tau$. When $\tau=0$, neither anti-phase nor in-phase stable oscillations can be excited even for very strong coupling. Apparently, the shortest time delay is determined by a minimal duration of a spike impulse in the neuron under excitation by the delayed coupling. In return, the threshold with respect to $\sigma$ is caused by a minimal amplitude of the influence of the neighboring neuron through the delayed coupling. The smaller the delay, the stronger the coupling strength must be, up to a certain value of $\tau$, after that the threshold with respect to $\sigma$ does not change. It should be noted that, as is clearly seen from Fig.~\ref{fig:2neuron_LE} and is verified by our calculations, the $\sigma$ threshold is substantially lower for larger values of $\gamma$. Increasing $\gamma$ decreases the neuron dissipation, and thus, a lower amplitude of excitation is needed to induce oscillations in the FitzHugh--Nagumo neuron. The diagrams of the maximal Lyapunov exponent in Fig.~\ref{fig:2neuron_LE} give evidence that the threshold for the oscillatory regime with respect to the delay is larger for in-phase oscillations than that for anti-phase ones, while the $\sigma$ threshold is practically the same for both cases. Note that negative values of the maximal Lyapunov exponent in the quiescent region decrease with an increase in a duration of the delay $\tau$.

\subsection{Linear stability analysis of the equilibrium}\label{secIIIB}
Let the unique fixed point be $\textbf{x}^{*}= (x_{1}^{*}, y_{1}^{*}, x_{2}^{*}, y_{2}^{*})$. The fixed point is given by solving simultaneously two nonlinear equations for each pair $(x_{1}^{*}, y_{1}^{*})=(x_{2}^{*}, y_{2}^{*})$:
\begin{equation}
\begin{aligned}
    x_{1} - \frac{x_{1}^{3}}{3} - y_{1} &= 0,\\
    \gamma x_{1} - y_{1} + \beta &= 0,\\
    x_{2} - \frac{x_{2}^{3}}{3} - y_{2} &= 0,\\
    \gamma x_{2} - y_{2} + \beta &= 0
\end{aligned}
\label{eq:CoupleddElayFHN_eqn}
\end{equation}

\begin{figure}[!t]
\centering
$\gamma=0.5$ \hspace{93pt} $\gamma=0.7$
\\
\includegraphics[width=0.495\linewidth]{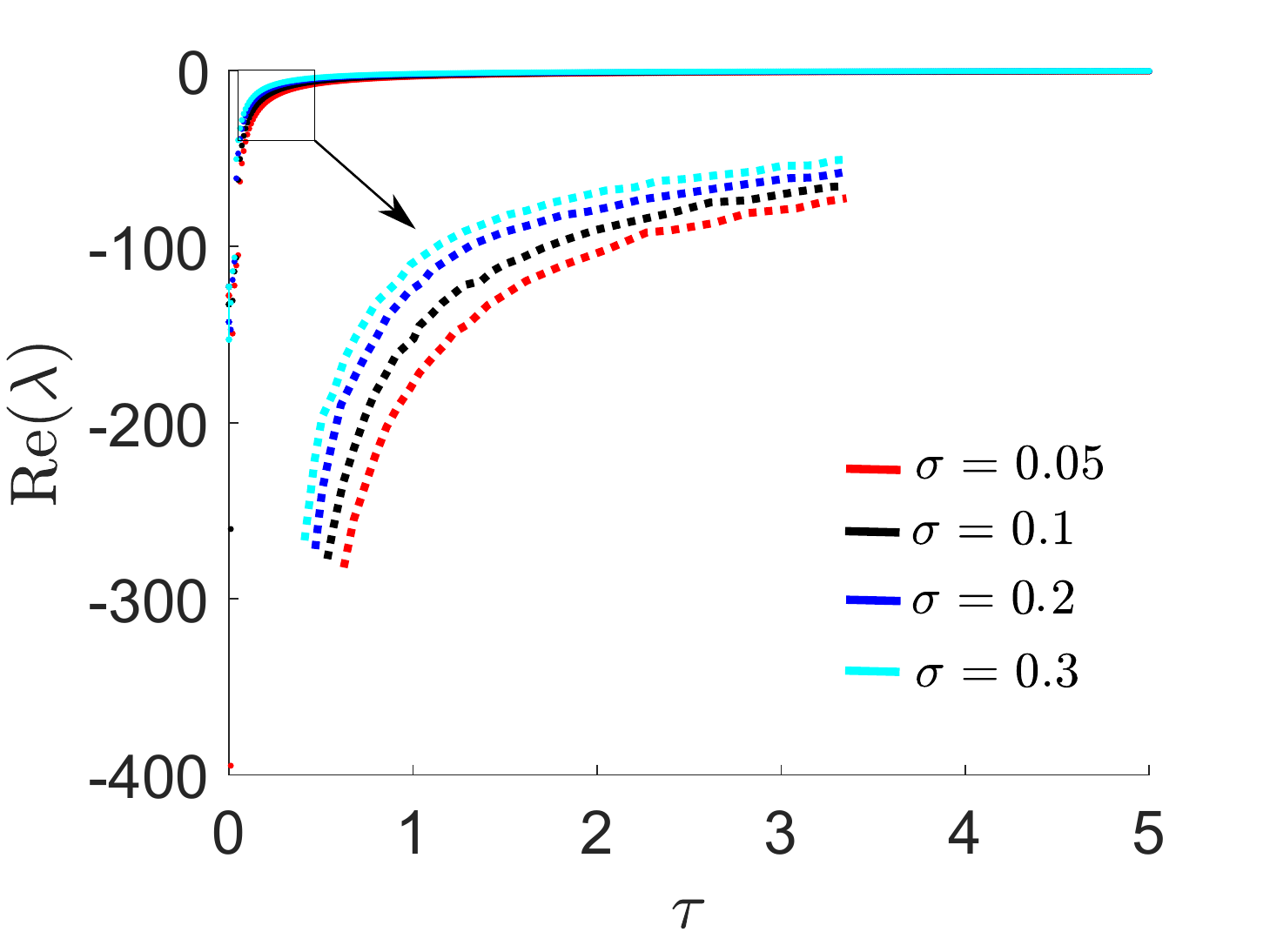}
\includegraphics[width=0.495\linewidth]{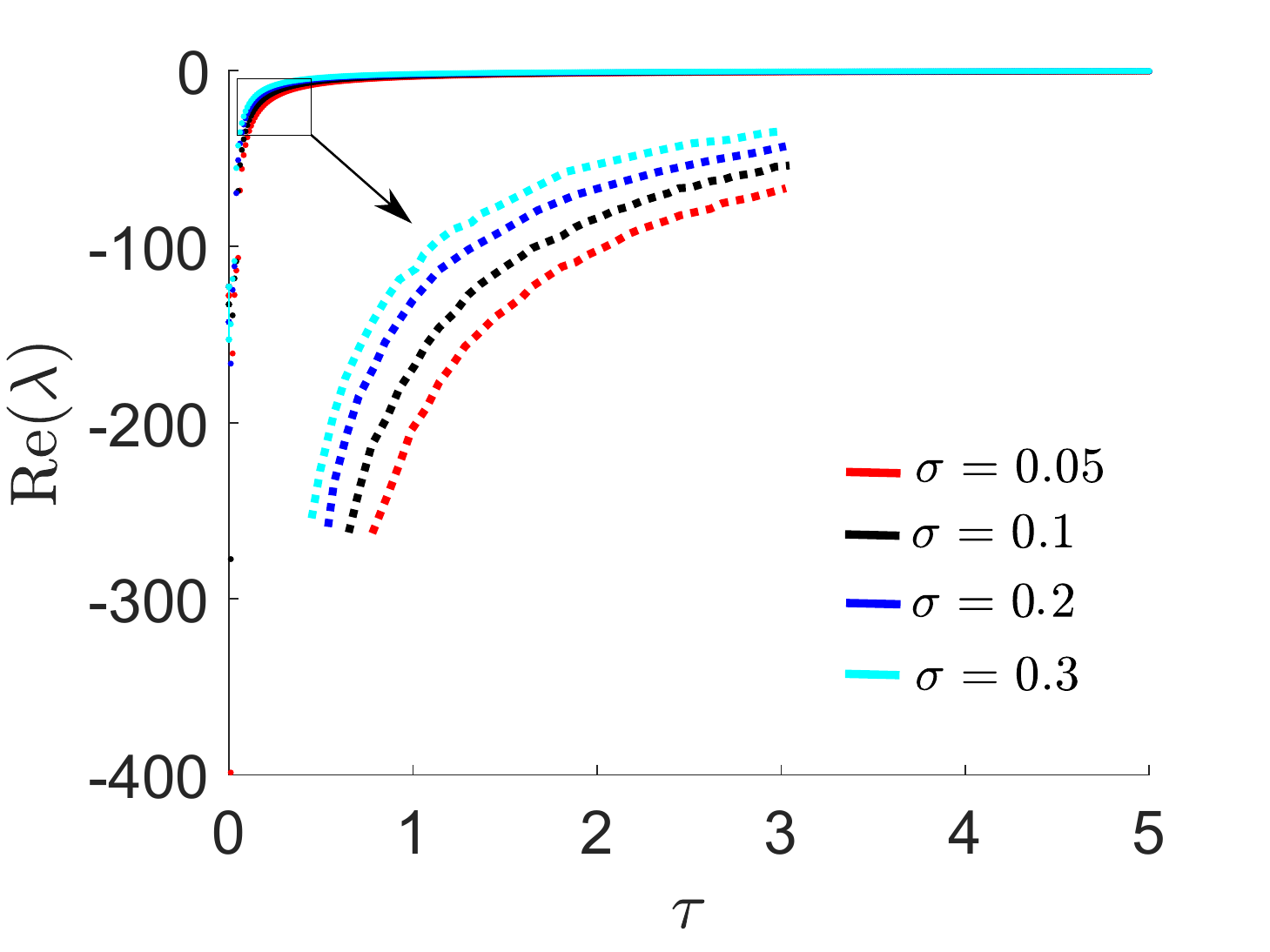}
\\
(a)\hspace{113pt}(b)
\caption{Real part of eigenvalues $\lambda$ versus the delay time $\tau$ for $\gamma=0.5$ (a) and $\gamma=0.7$ (b). The equilibrium point of the system~\eqref{eq:CoupleddElayFHN_eqn} for two coupled FitzHugh--Nagumo units is stable for all parameter values shown. The curve shifts slightly upwards with increasing $\sigma$ (see blow-up in the insets). The parameters are set as $\varepsilon=0.01, \beta=-0.5$.}
\label{fig:TwoCoupledStabilityAnalysis}
\end{figure}

\begin{figure}[!t]
\centering
\includegraphics[width=0.495\linewidth]{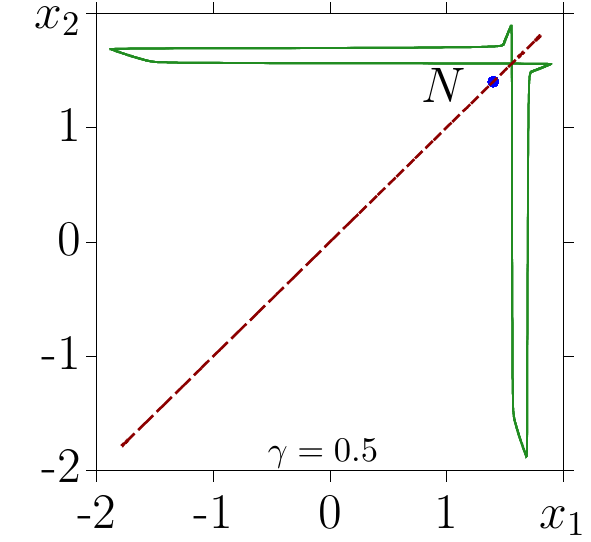}
\includegraphics[width=0.495\linewidth]{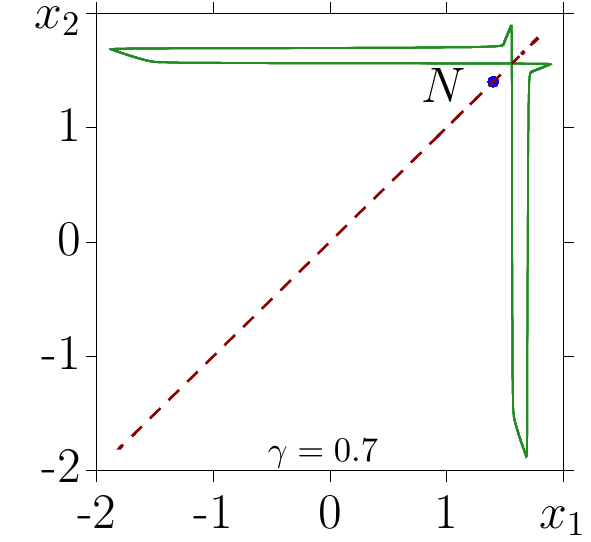}
\\
(a)\hspace{113pt}(b)
\caption{Phase projections in the ($x_1,x_2$) coordinate plane for the system~\eqref{eq:twoFHN} at $\gamma=0.5$ (a) and $\gamma=0.7$ (b) with $\varepsilon = 0.01$ and $\beta=-0.5$. In the planes the stable equilibrium $E$, the in-phase  (dotted red line) and the anti-phase  (green curve)  periodic orbits are shown.}
\label{fig:FHNmultistability}
\end{figure}

The equilibrium point in the excitable regime is given by $x_{1}^{*} = x_{2}^{*}$ and $y_{1}^{*} = y_{2}^{*}$. Linearizing system~\eqref{eq:twoFHN} around the fixed point $\textbf{x}^{*}$ by setting $\textbf{x}(t) = \textbf{x}^{*} + \delta \textbf{x}(t)$, one obtains
\begin{equation}
    \delta \dot{\mathbf{x}} = \frac{1}{\varepsilon} A \delta \mathbf{x}(t) + \frac{\sigma}{\varepsilon} B \delta \mathbf{x}(t-\tau),
\end{equation}
where
\begin{align}
A &= \begin{bmatrix}
 \xi & -1 & 0 & 0\\
 \gamma \varepsilon & -\varepsilon & 0 & 0\\
 0 & 0 & \xi & -1\\
 0 & 0& \gamma \varepsilon & -\varepsilon
\end{bmatrix},
\\
B &= \begin{bmatrix}
 0 & 0 & 1 & 0\\
 0 & 0 & 0 & 0\\
 1 & 0 & 0 & 0\\
 0 & 0 & 0 & 0
\end{bmatrix},
\end{align}
where we have introduced $\xi := 1-x_{1}^{*^{2}} - \sigma = 1-x_{2}^{*^{2}} - \sigma < 0$ since ${x_1^*}^2>1$. The ansatz 
\begin{equation}
    \delta \mathbf{x}(t) =  \exp(\lambda t)\mathbf{u},
\end{equation}
where $\mathbf{u}$ is an eigenvector of the Jacobian matrix $A$, leads to the following characteristic equation for the eigenvalues $\lambda$, given by setting the determinant 
$$
\rm{det}\Big(\frac{A}{\varepsilon} + \frac{\sigma}{\varepsilon} B \exp(-\lambda \tau) - \lambda I_{N \times N} \Big) = 0.
$$
Expanding the matrices inside the determinant, we get
\begin{equation}
\begin{vmatrix}
\frac{\xi}{\varepsilon} - \lambda & -\frac{1}{\varepsilon} & \sigma e^{-\lambda \tau} & 0\\
\gamma & -1 - \lambda & 0 & 0\\
\sigma e^{-\lambda \tau} & 0 & \frac{\xi}{\varepsilon} - \lambda & -\frac{1}{\varepsilon}\\
0 & 0 & \gamma & -1-\lambda
\end{vmatrix} = 0.
\end{equation}
%which gives
%\begin{equation*}
%\begin{aligned}
%\Bigl( \frac{\xi}{\varepsilon}-\lambda\Bigr) (-1-\lambda) \Bigl[ \Bigl(\frac{\xi}{\varepsilon} - \lambda \Bigr) (-1-\lambda) + \frac{\gamma}{\varepsilon}\Bigr] & +
%\\
%+ \frac{\gamma}{\varepsilon} \Bigl[ \Bigl(\frac{\xi}{\varepsilon} - \lambda\Bigr) (-1-\lambda) + \frac{\gamma}{\varepsilon} \Bigr] & +
%\\ + \sigma e^{-\lambda \tau} (1+\lambda) \sigma e^{-\lambda \tau} (-1-\lambda) & = 0,
%\end{aligned}
%\end{equation*}
Because of the symmetry between 1 and 2 the determinant can be factorized in analogy with the simplified model~\cite{Scholl2009} into
\begin{equation}
\Bigl[ \Bigl( \frac{\xi}{\varepsilon} - \lambda\Bigr)(-1-\lambda) + \frac{\gamma}{\varepsilon}\Bigr]^2 - \Bigl( \frac{\sigma}{\varepsilon} e^{-\lambda \tau} (-1-\lambda)\Bigr)^2 = 0,
\end{equation}
which gives the characteristic equation
\begin{equation}
    \lambda^2 - \lambda\Bigl(\frac{\xi}{\epsilon} - 1\Bigr) - \frac{\xi}{\epsilon} + \frac{\gamma}{\epsilon} \pm (\lambda+1) \frac{\sigma}{\varepsilon} e^{-\lambda \tau} = 0.
    \label{eq:charac}
\end{equation}
This transcendental equation has infinitely many complex solutions $\lambda$. Fig. \ref{fig:TwoCoupledStabilityAnalysis} shows the largest real part of $\lambda$ versus $\tau$.
Obtaining $\xi$ from the above equation does not admit a simple form as with~\cite{Scholl2009}.
%As $\sigma$ increases, the curve where the equilibrium point is stable (with real parts of the eigenvalue negative) shifts upwards, but the shape of the curve is almost similar for various values of $\sigma$.
For all values of $\tau$, the real part is negative and the modulus of the real part of the eigenvalue decreases monotonically as $\tau$ increases.
Asymptotically for large $\tau$ the eigenvalues are given by the quadratic characteristic equation
$$\lambda^{2} + \lambda\Bigl(\frac{|\xi|}{\varepsilon}+1\Bigr) + \frac{|\xi|+\gamma}{\varepsilon} = 0.$$
where we have used that $\xi<0$. Its solutions
\begin{equation}
    \lambda_{1,2} = -\frac{1}{2} \Bigl( \frac{|\xi|}{\varepsilon} + 1 \Bigr) \Bigl( 1 \mp \sqrt{1 - 4\varepsilon \Bigl(\frac{|\xi| + \gamma}{(|\xi| + \varepsilon)^{2}}\Bigr)} \Bigr).
    \label{eq:rootlambda}
\end{equation}
have always negative real parts. Hence the equilibrium is always stable. For both $\gamma = 0.5$ (in Fig. \ref{fig:TwoCoupledStabilityAnalysis},(a)), and $\gamma = 0.7$ (in Fig. \ref{fig:TwoCoupledStabilityAnalysis},(b)), we plot ${\rm{Re}}(\lambda)$ versus $\tau$. Increasing $\gamma$ does not seem to affect the stability of equilibrium point with respect to $\tau$. Thus, the equilibrium point coexists with limit cycle oscillations, and there is multistability in the two coupled FitzHugh--Nagumo neurons. This property is verified numerically and illustrated in Fig.~\ref{fig:FHNmultistability} for two values of the parameter $\gamma$. It is seen that there are three stable regimes in the phase plane, i.e., the equilibrium point $N$ and two limit cycles, one corresponding to in-phase oscillations (dotted red line) and the other one to anti-phase oscillations (green curve).

%%%%%%%%%%%%%%%%%%%%%%%%%%%%%%% fig.7 %%%%%%%%%%%%%%%%%%%%%%%%%%%%%%%%%
%%%%%%%%%%%%%%%%%%%%%%%%%%%%%%%%%%%%%%%%%%%%%%%%%%%%%%%%%%%%%%%%%%%%%%%%%

%%%%%%%%%%%%%%%%%%%%%%%%%%%%%%% fig.7b %%%%%%%%%%%%%%%%%%%%%%%%%%%%%%%%%
%%%%%%%%%%%%%%%%%%%%%%%%%%%%%%%%%%%%%%%%%%%%%%%%%%%%%%%%%%%%%%%%%%%%%%%%%

\section{Dynamics of a ring of FitzHugh--Nagumo neurons with time-delayed coupling}

Knowing the features of emergence of oscillations in the two FitzHugh--Nagumo neurons with delayed coupling, we extend our studies to a ring of excitable FitzHugh--Nagumo neurons with delayed coupling. This network is described by the following system of equations:
%%%%%%%%%%%
\begin{equation}
\begin{array}{l}
\begin{aligned}
\varepsilon \dfrac{dx_i(t)}{dt} = x_i(t) -  x_i^3(t)/3 -y_i(t) + \dfrac{\sigma}{2P}\sum_{j=i-P}^{i+P} [x_{j}(t-\tau) - \\- x_{i}(t)],
\end{aligned}
\\[8pt]
\dfrac{dy_i(t)}{dt} = \gamma x_i(t) - y_i(t) + \beta,\\[8pt]
i=1,\ldots,N,~~x_{i+N}(t)=x_{i}(t),~y_{i+N}(t)=y_{i}(t),
\end{array}
\label{eq:ring}
\end{equation}
%%%%%%%
where the index $i$ of the dynamical variables $x_i$ and $y_i$, $i = 1,\ldots, N$ determines the position in the network, and $N=50$ is the total number of elements. The parameter $P$ is the coupling range and denotes the number of neighbouring nodes which the $i$th oscillator is coupled with from each side. Here we consider the case when all the oscillators are identical and each of them interacts (through the variable $x$) with one node from the left and one from the right. Thus, $P=1$ corresponds to local coupling between the nodes. The boundary conditions are periodic, and the initial states are fixed for each neuron as follows: $x_{i}(t\in[-\tau;0])=x_{i}^0$ and $y_{i}(t\in[-\tau;0])=y_{i}^0$, where $x_i^0$ and $y_i^0$ are randomly uniformly distributed within the intervals: $x_i(0) \in [-2,2], y_i(0) \in [-1,1]$. The other parameters have the same meaning as for the system \eqref{eq:twoFHN}. Delay-coupled networks of simplified FitzHugh--Nagumo systems have been studied numerically and analytically for a ring and for more general topologies in Refs.\cite{Nikitin2019, Plotnikov2016, Lenhert2011}.

\subsection{Delay-induced oscillations}

We vary the coupling strength $\sigma$ and the  delay time $\tau$ in the coupling term as in the case of the two coupled FitzHugh--Nagumo neurons described above. Interestingly, the delay-induced oscillations in the ring Eq.~\eqref{eq:ring} can be only in-phase synchronized for any initial conditions under consideration, and stable anti-phase oscillations are not observed.  

\begin{figure}[!t]
\centering
$\gamma=0.5$ \hspace{93pt} $\gamma=0.7$
\\
\includegraphics[width=0.495\linewidth]{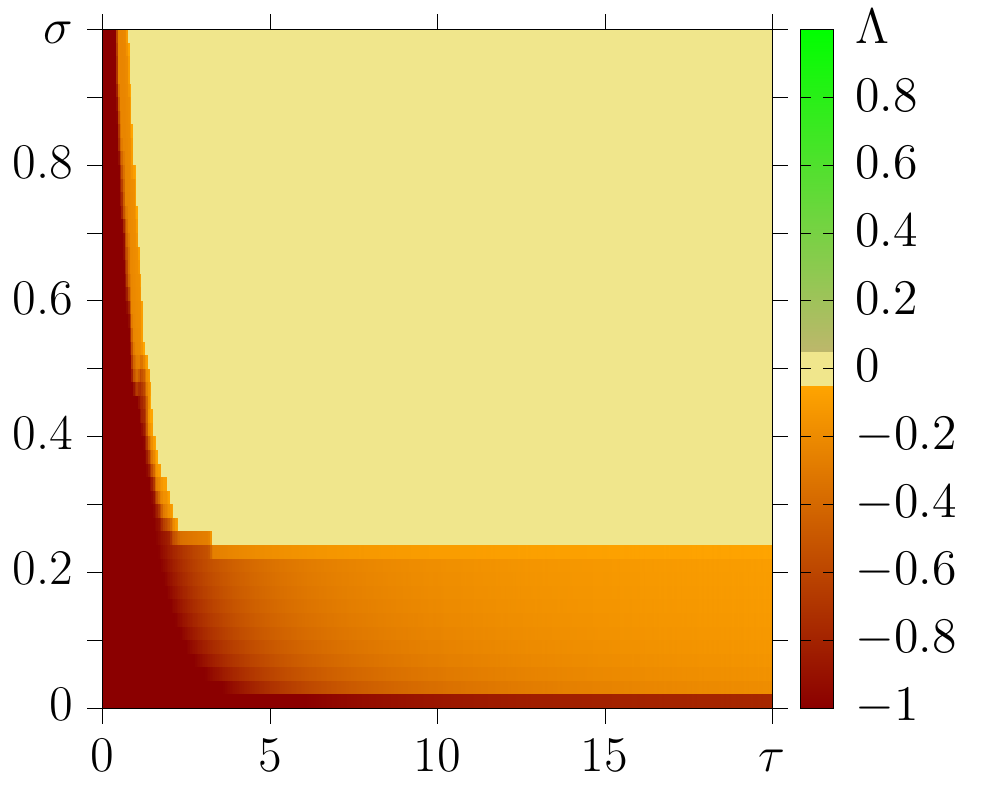}
\includegraphics[width=0.495\linewidth]{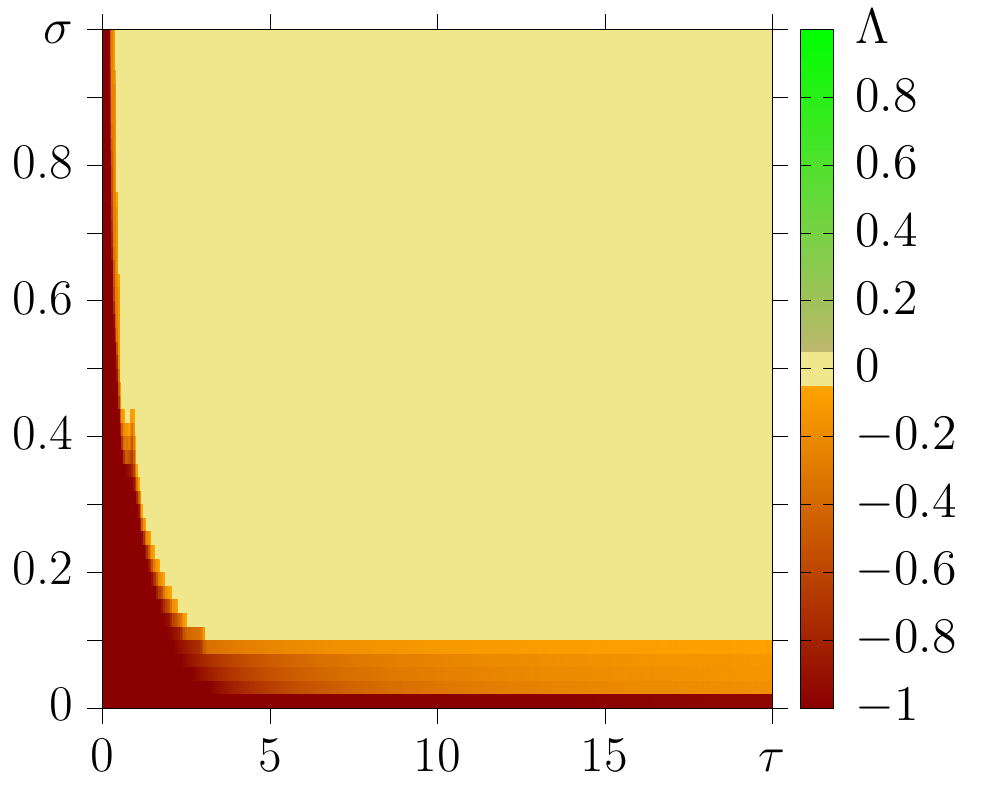}
\\
(a)\hspace{113pt}(b)
\caption{Distributions of the maximal Lyapunov exponent $\Lambda$ in the ($\tau,\sigma$) parameter plane for two different values of the dissipation parameter $\gamma=0.5$ (a), and $\gamma=0.7$ (b) for random initial conditions in the network~\eqref{eq:ring}. Other parameters: $\varepsilon=0.01$, $\beta=-0.5$, $P=1$, $N=50$.}
\label{fig:ring_regimes}
\end{figure}

We start with analyzing how the threshold of the oscillation excitation depends on the coupling strength and the delay. We also calculate distributions of the maximal Lyapunov exponent $\Lambda$ in  the ($\tau,\sigma$) parameter plane, which are presented in Fig.~\ref{fig:ring_regimes} for two different values of the dissipation parameter $\gamma$.
%%%%%%%%%%%%%%%%%%%%%%%%%%%%%%% fig.8 %%%%%%%%%%%%%%%%%%%%%%%%%%%%%%%%%
%%%%%%%%%%%%%%%%%%%%%%%%%%%%%%%%%%%%%%%%%%%%%%%%%%%%%%%%%%%%%%%%%%%%%%%%%

Our simulations show that all delay-induced oscillations in the neural network~\eqref{eq:twoFHN} are regular ($\Lambda=0$ within the whole region of oscillation existence (Fig.~\ref{fig:ring_regimes})).
Comparing the diagrams constructed for the two delay-coupled FitzHugh--Nagumo neurons (Fig.~\ref{fig:2neuron_LE}) and the neural network (Fig.~\ref{fig:ring_regimes}) with delayed coupling gives evidence of a similarity in the emergence of oscillations with respect to the coupling strength and the delay in both cases. The maximal Lyapunov exponent $\Lambda$ decreases with decreasing delay in the non-oscillatory regime as in the case of two coupled neurons. However, the threshold with respect to $\sigma$ is slightly larger for the network.
% The analytical methods used for the simpilied form of the equations ... in [...] can not be applied to estimate analytically the parameter values in this case. Due to compexity of the linear stability analysis we only use the numerical results [...].}
Thus, compared to the two coupled neurons, the region of self-sustained oscillations decreases both with respect to the dissipation parameter $\gamma$ and delay $\tau$. Since the scenario of oscillation excitation is very similar for $\gamma=0.5$ and $\gamma=0.7$, we continue our numerical studies of the neural network dynamics for the case of $\gamma=0.5$.

\begin{figure}[!h]
\centering
transient process \hspace{15pt} asymptotic process \hspace{18pt} snapshot $x_i(t)$
\\
\includegraphics[width=0.32\linewidth]{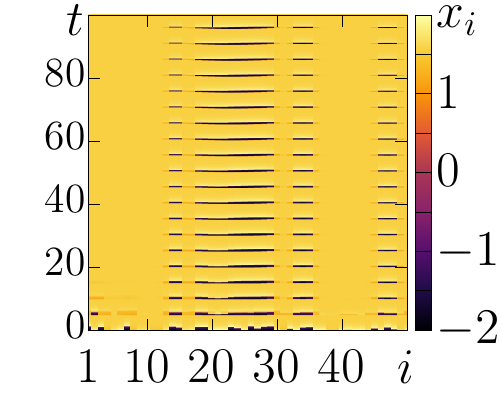}
\includegraphics[width=0.32\linewidth]{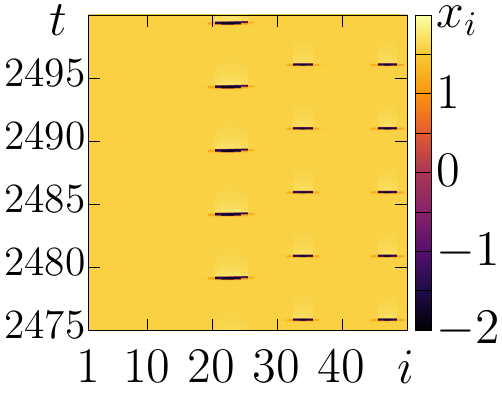}
\includegraphics[width=0.32\linewidth]{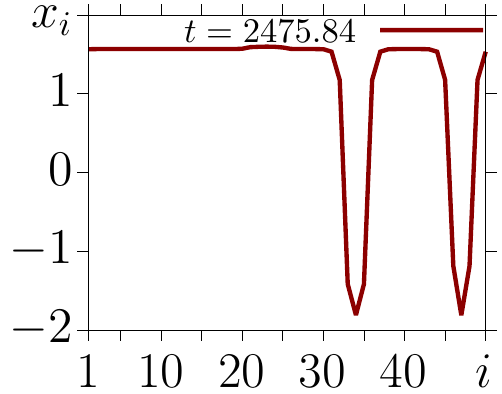}
\\
$\sigma=0.3\;\;\;$(a1)\hspace{67pt}(a2)\hspace{67pt}(a3)$\qquad$
\\
\includegraphics[width=0.32\linewidth]{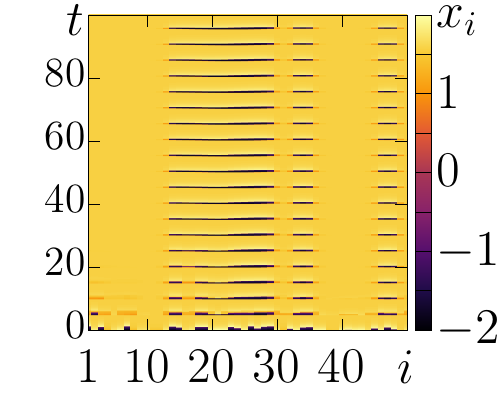}
\includegraphics[width=0.32\linewidth]{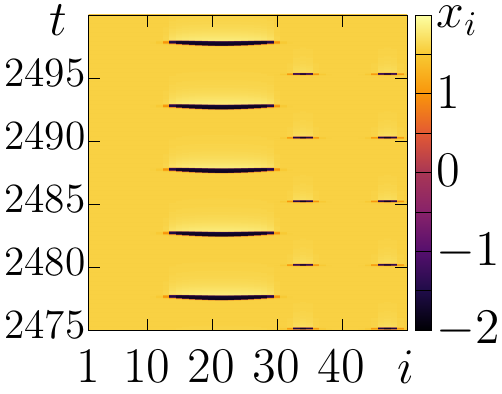}
\includegraphics[width=0.32\linewidth]{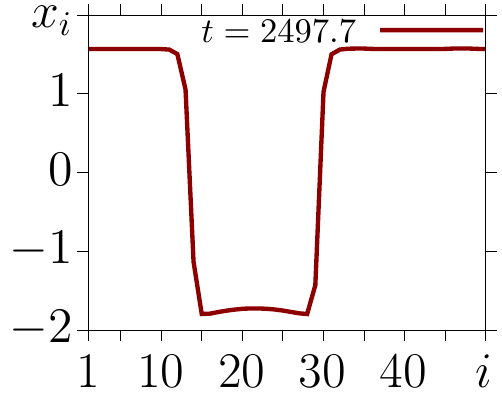}
\\
$\sigma=0.4\;\;\;$(b1)\hspace{67pt}(b2)\hspace{67pt}(b3)$\qquad$
\\
\includegraphics[width=0.32\linewidth]{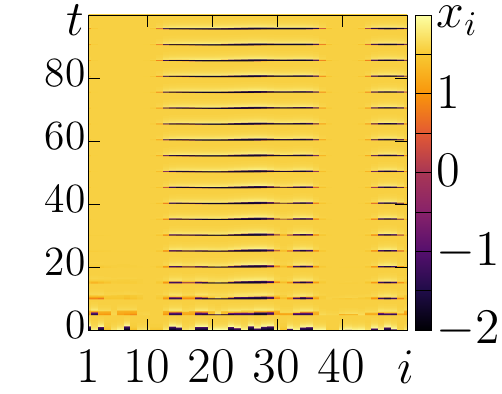}
\includegraphics[width=0.32\linewidth]{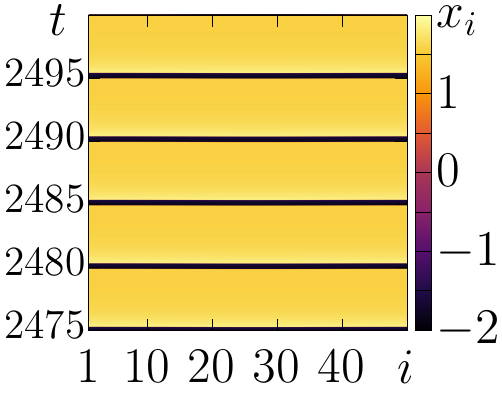}
\includegraphics[width=0.32\linewidth]{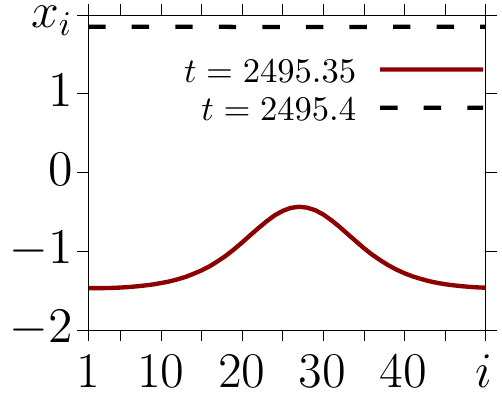}
\\
$\sigma=0.5\;\;\;$(c1)\hspace{67pt}(c2)\hspace{67pt}(c3)$\qquad$
\caption{Delay-induced oscillations observed within the region with $\Lambda\approx 0$ in the diagram in Fig.~\ref{fig:ring_regimes},(a) for three increasing values of the coupling strength: $\sigma=0.3$ (a1)-(a3), $\sigma=0.4$ (b1)-(b3), and $\sigma=0.5$ (c1)-(c3). The evolution of the network \eqref{eq:ring} dynamics is illustrated by space-time plots for the transient process within $t\in[0,250]$ (left column, (a1)-(c1)) and asymptotic process (middle column, (a2)-(c2)) within $t\in[2475,2500]$, and by snapshots for the $x_i$ variables (right column, (a3)-(c3), red curves). The dotted line in (c3) corresponds to the quiescent regime. Note that the time scales in the left and middle columns are different. Other parameters: $\tau=5$, $\gamma=0.5$, $\varepsilon=0.01$, $\beta=-0.5$, $P=1$, $N=50$.}
\label{fig:ring_STP}
\end{figure}

We now fix the delay time to $\tau=5$ and explore the evolution of the ring dynamics as the coupling strength $\sigma$ increases. When $\sigma$ is sufficiently small, the initial fluctuations cannot induce oscillations in the network \eqref{eq:ring} (red and orange colors in the distribution in Fig.~\ref{fig:ring_regimes}). Delay-induced oscillations occur when $\Lambda\approx 0$ (light-yellow colors  in the diagrams in Fig.~\ref{fig:ring_regimes}). However, unlike the case of two coupled neurons, only a part of the network begins to demonstrate oscillations, while the rest remains quiescent. This is similar to a bump state in neuroscience\cite{laing2020moving, schmidt2020bumps}. The transient process for $\sigma=0.3$ is illustrated by a space-time plot in Fig.~\ref{fig:ring_STP},(a1). It is seen that the transient process lasts a sufficiently long time within several dozens of periods. There are one wide and three narrow clusters which demonstrate  the oscillatory dynamics at the initial stage. However, in the course of time, the size of the clusters changes and the wide cluster significantly narrows and one of the small clusters completely disappears. As a result, only three clusters are observed asymptotically, which is shown in the space-time plot in Fig.~\ref{fig:ring_STP},(a2), where the vertical time scale is expanded for better visualization. Note that the neurons in the two clusters are excited in-phase with each other, while their instantaneous phases do not coincide with those for the neurons in the third cluster. Apparently, the oscillation period is not strictly equal to the time delay, $T=\tau$, but is very close to it, $T\approx\tau$. Furthermore, the period can slightly change in different clusters independently from each other. This leads to an advance or delay of the instantaneous phases over a long observation time. The rest of the ring oscillators remain quiescent. Moreover, the oscillators of the clusters with oscillating dynamics demonstrate their activity during a sufficiently short time and are at rest most of the time between pulses caused by the delay, which is characteristic for the FitzHugh--Nagumo oscillator dynamics. The instantaneous spatial profile at the moment of oscillation excitation is shown in Fig.~\ref{fig:ring_STP},(a3). The neurons are not excited simultaneously but with a certain time lag. The neurons belonging to the quiescent cluster and interacting with the edge elements of the oscillatory clusters are also perturbed during the pulse excitation but this perturbation is still insufficient to excite the neurons. Apparently, increasing the coupling strength $\sigma$ can lead to the excitation of other neurons.

%%%%%%%%%%%%%%%%%%%%%%%%%%%%%%% fig.9 %%%%%%%%%%%%%%%%%%%%%%%%%%%%%%%%%
%%%%%%%%%%%%%%%%%%%%%%%%%%%%%%%%%%%%%%%%%%%%%%%%%%%%%%%%%%%%%%%%%%%%%%%%%

We increase the coupling strength to $\sigma=0.4$. The corresponding space-time plots and the snapshot of the ring dynamics are presented in Fig.~\ref{fig:ring_STP},(b1)-(b3). As can be seen from the plots, increasing $\sigma$ causes the change in the transient time of the process. The initial spatial distribution of the clusters is very similar to the case of $\sigma=0.3$ (Fig.~\ref{fig:ring_STP},(a1)). However, the sizes of the oscillatory clusters do not practically change in time and the transient process becomes sufficiently short. At the same time, the oscillation period in different clusters continues to change slightly around the delay time $\tau$ and the neurons are not activated simultaneously. The stable regime is exemplified by the space-time plot pictured in Fig.~\ref{fig:ring_STP},(b2). The width of the oscillatory cluster remains the same as at the initial stage of oscillations. The instantaneous phases of oscillations in different clusters are shifted with respect to each other, and their phase differences are not constant in time, as in the case of $\sigma=0.3$. The snapshot of the system state taken for the stable regime at the moment of oscillation excitation (Fig.~\ref{fig:ring_STP},(b3)) gives evidence that the neurons are also excited with a certain lag as in the previous case.

We continue to increase the coupling strength between the FitzHugh--Nagumo neurons in~\eqref{eq:ring} and now consider the case of $\sigma=0.5$. As follows from the space-time plot for the transient process presented in Fig.~\ref{fig:ring_STP},(c1), the oscillations originate in the same way as in the two previously considered cases due to the same set of initial conditions. However, all the oscillatory clusters expand over time. This effect is related to the fact that the edge neurons of the oscillatory clusters interact with the quiescent neurons. Since the coupling strength is sufficiently large, this interaction excites the quiescent neurons and thus, they join to the oscillatory clusters. In turn, these neurons subsequently excite the adjacent elements. Thus, all the neurons of the network~\eqref{eq:ring} begin to demonstrate the oscillatory dynamics which is observed over a sufficiently long time. This asymptotically stable regime is illustrated by the space-time plot shown in Fig.~\ref{fig:ring_STP},(c2). The neurons are not excited simultaneously, there is a short time lag at the initial state of excitation. The oscillation period is strictly equal to $T=\tau$. The instantaneous phases of all the nodes are close to each other but not equal, which shows a snapshot of the system state in Fig.~\ref{fig:ring_STP},(c3)). The instantaneous profile shows a certain spatial distribution of the $x_i$ values at the moment of neuron excitation.

%%%%%%%%%%%%%%%%%%%%%%%%%%%%%%% fig.5 %%%%%%%%%%%%%%%%%%%%%%%%%%%%%%%%%
\begin{figure}[!t]
\centering
\includegraphics[width=0.243\linewidth]{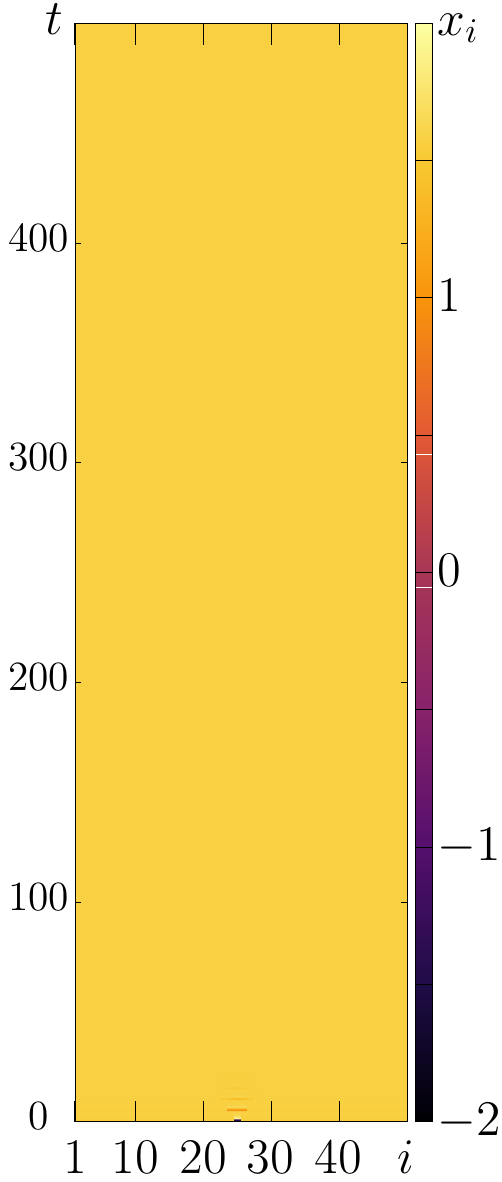}
\includegraphics[width=0.243\linewidth]{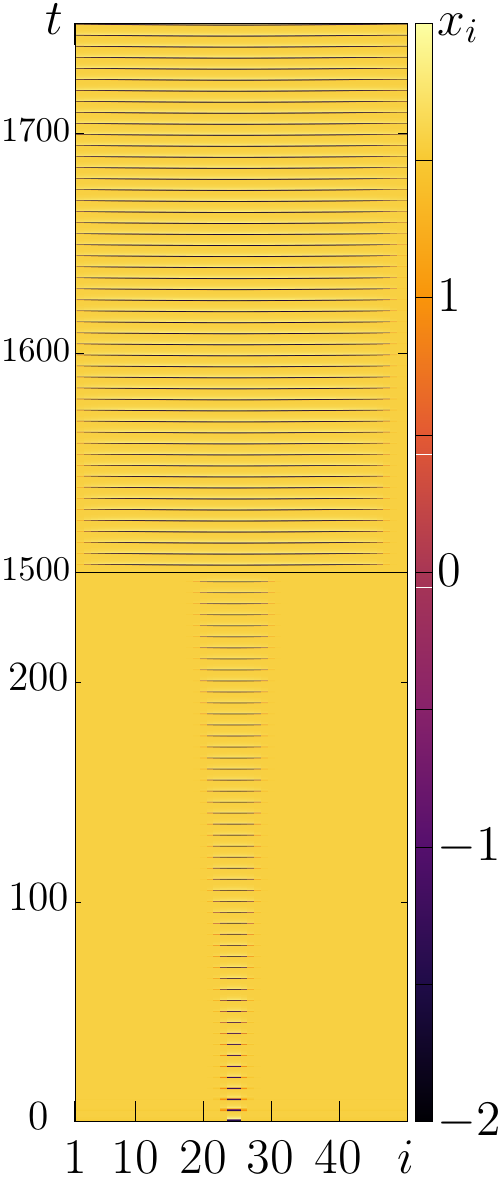}
\includegraphics[width=0.243\linewidth]{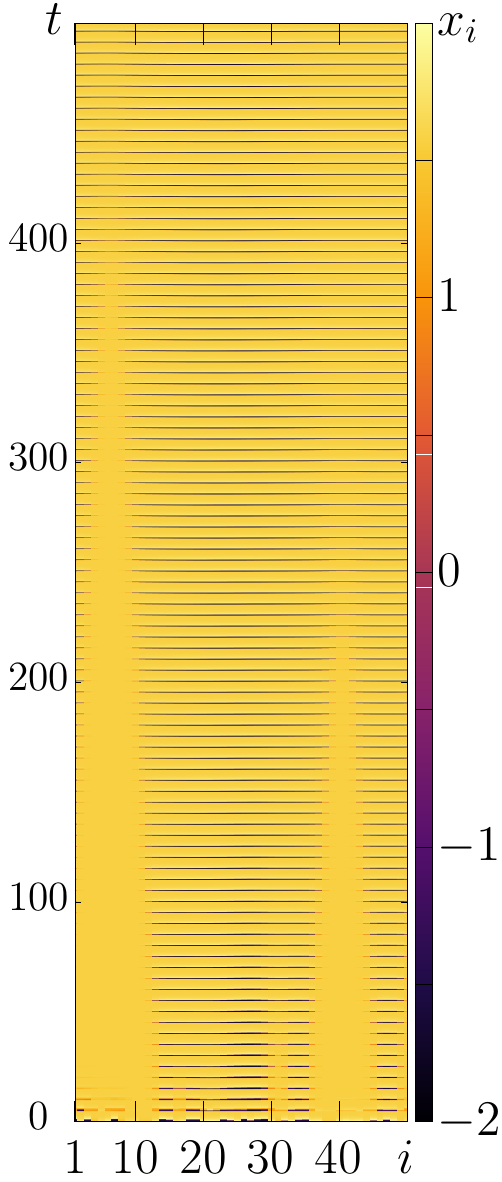}
\includegraphics[width=0.243\linewidth]{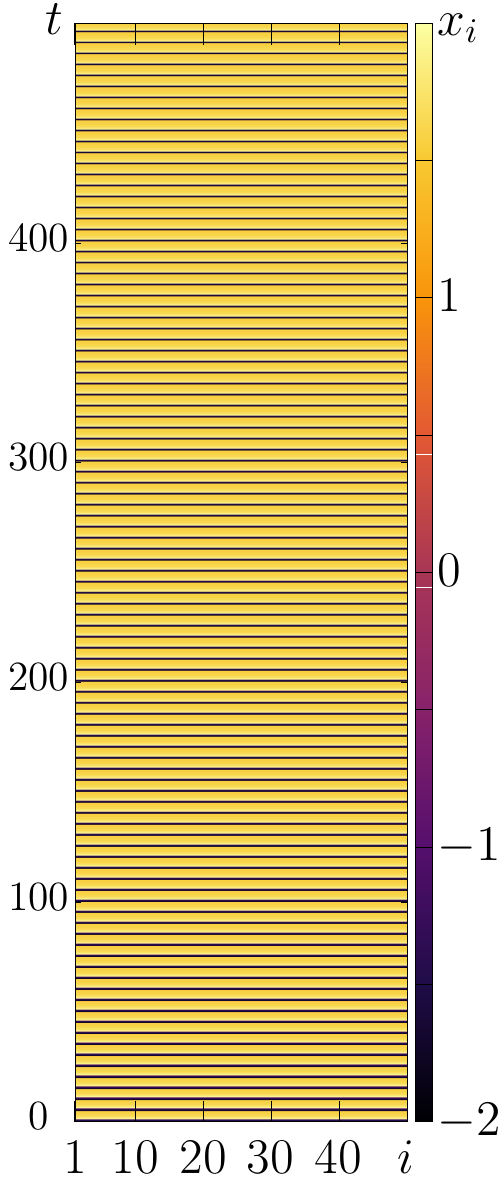}
\\
(a)\hspace{47pt}(b)\hspace{47pt}(c)\hspace{47pt}(d)
\caption{Delay-induced oscillations observed within the region with $\Lambda\approx 0$ in the diagram in Fig.~\ref{fig:ring_regimes},(a) for one (a), two (b), random (about 30 nodes) (c) and 50 (d) initially excited neurons with $\sigma=0.5$. The evolution of the network \eqref{eq:ring} dynamics is illustrated by space-time plots. Other parameters: $\tau=5$, $\gamma=0.5$, $\varepsilon=0.01$, $\beta=-0.5$, $P=1$, $N=50$.}
\label{fig:ringICsets}
\end{figure}
%%%%%%%%%%%%%%%%%%%%%%%%%%%%%%%%%%%%%%%%%%%%%%%%%%%%%%%%%%%%%%%%%%%%%%%%%

Our numerical simulations show that the duration of the transient process can depend on the number of initially excited neurons. This fact is illustrated in Fig.~\ref{fig:ringICsets} where the space-time plots are depicted for the cases when  one, two, random (about 30 nodes), and all the network neurons ($N=50$) are  initially excited. 
The numerical results testify that  there is a critical number of initially excited neighbouring neurons which allows the ring to demonstrate the oscillatory behavior. All the network neurons remain in the quiescent state if only a single  neuron is initially excited (Fig.~\ref{fig:ringICsets}(a)) and show a gradual propagation of oscillation activity when at least two neighbouring neurons are initially excited (Fig.~\ref{fig:ringICsets}(b)). However, in this case the transient process is rather long, as is seen in Fig.~\ref{fig:ringICsets}(b), while all the neurons are excited more rapidly when randomly chosen neighbouring  neurons or all the neurons are initially excited (Fig.~\ref{fig:ringICsets}(c,d)). 

\begin{figure}[!h]
\centering
\includegraphics[width=0.32\linewidth]{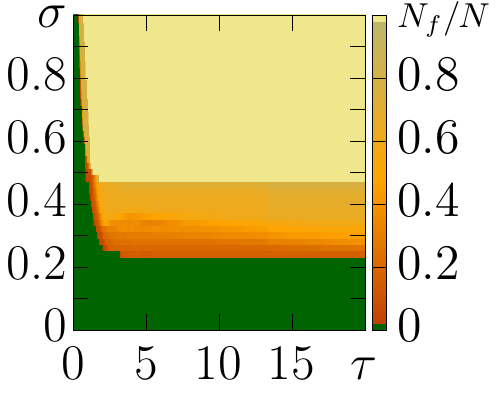}
\includegraphics[width=0.32\linewidth]{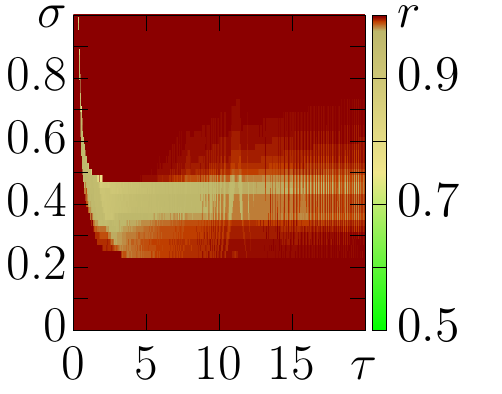}
\includegraphics[width=0.32\linewidth]{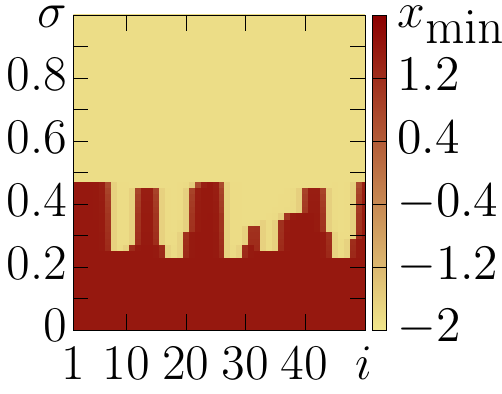}
\\
(a)\hspace{72pt}(b)\hspace{72pt}(c)
\\
\includegraphics[width=0.32\linewidth]{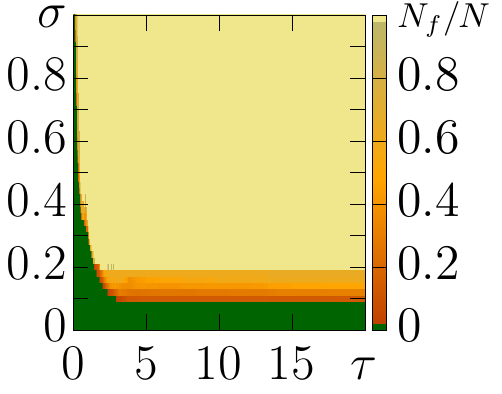}
\includegraphics[width=0.32\linewidth]{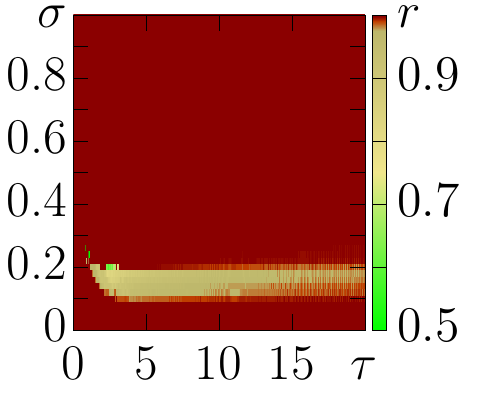}
\includegraphics[width=0.32\linewidth]{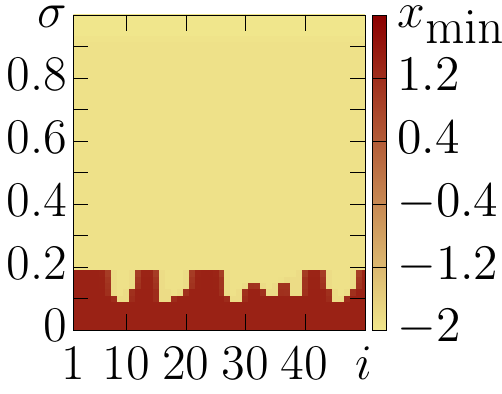}
\\
(d)\hspace{72pt}(e)\hspace{72pt}(f)
\caption{Diagrams for the ratio of the number of firing nodes $N_f$ to the whole number of neurons (a,d) and for the global order parameter $r$ (b,e) in the $(\tau,\sigma)$ parameter plane, and spatial distributions of minimal values of variable $x$ (quiescent: red, oscillatory: yellow) (c,f) for the delay time $\tau=5$ in the network~\eqref{eq:ring} for two different values of the dissipation parameter $\gamma=0.5$ (a-c) and $\gamma=0.7$ (d-f). Other parameters: $\varepsilon=0.01$, $\beta=-0.5$,  $P=1$, $N=50$.}
\label{fig:ring_ratio}
\end{figure}

In order to demonstrate how the coupling strength $\sigma$ affects the number of excited (firing) neurons, we calculate distributions of the ratio of the number of firing nodes $N_f$ to the whole number of neurons in the network $N$ (Fig.~\ref{fig:ring_ratio},(a,d)), the global order parameter
\[
\begin{aligned}
r & = \left\langle \left| \frac{1}{N} \sum_i e^{i\Theta_i} \right| \right\rangle_t, \\
%r=\left\langle \sqrt{ \left\langle \dfrac{x_i}{\sqrt{x_i^2+y_i^2}} \right\rangle_{i}^2 + \left\langle \dfrac{y_i}{\sqrt{x_i^2+y_i^2}} \right\rangle_{i}^2 } \right\rangle_{t}
\end{aligned}
\]
where $\Theta_i= \arctan(y_i/x_i)$ denotes the geometric phase of the $i$-th unit~\cite{Schoell2016}, 
in the $(\tau, \sigma)$ parameter plane (Fig.~\ref{fig:ring_ratio},(b,e)), and  spatial distributions of minimal values of $x_i$ variables in the $(i,\sigma)$  plane (Fig.~\ref{fig:ring_ratio},(c,f)). The notation $\langle\dots\rangle_t$ means averaging over time. The calculated dependences are plotted in Fig.~\ref{fig:ring_ratio} for two values of the dissipation parameter $\gamma=0.5$ (Fig.~\ref{fig:ring_ratio},(a-c)) and $\gamma=0.7$ (Fig.~\ref{fig:ring_ratio},(d-f)). The distributions $N_f/N(\tau,\sigma)$ are obtained using 10 different sets of random initial conditions for averaging the results.

 %%%%%%%%%%%%%%%%%%%%%%%%%%%%%%% fig.10 %%%%%%%%%%%%%%%%%%%%%%%%%%%%%%%%%
%%%%%%%%%%%%%%%%%%%%%%%%%%%%%%%%%%%%%%%%%%%%%%%%%%%%%%%%%%%%%%%%%%%%%%%%%

As follows from the diagrams shown in Fig.~\ref{fig:ring_ratio},(a,d) and (Fig.~\ref{fig:ring_ratio},(b,e)), two different threshold values with respect to the coupling strength $\sigma$ can be distinguished. Only a small part of the neurons are excited when the first threshold, $\sigma_{\rm th, min}$,  is exceeded. Below this value all the neurons are quiescent and the order parameter $r$ is very close to 1 (Fig.~\ref{fig:ring_ratio},(b,e)). As is seen from the diagrams, the first threshold value is rather large for the case of zero and very small delay times ($\tau <2$), rapidly decreases up to a certain level as the delay time increases ($\tau>4$) and then remains unchanged and independent of the delay time.  Note that this threshold value is higher for a larger dissipation (it is about $0.21$ for $\gamma=0.5$, Fig.~\ref{fig:ring_ratio},(a,b)) and lower for a smaller dissipation (it is about $0.1$ for  $\gamma=0.7$, Fig.~\ref{fig:ring_ratio},(d,e)).
%the threshold with respect to $\sigma$ and $\tau$ when at least one neuron begins to oscillate coincides with the region of zero values of the maximal Lyapunov exponent in the dependences depicted in Fig.~\ref{fig:ring_regimes}. 

When $\sigma$ exceeds the first threshold value $\sigma_{\rm th, min}$, the number of firing neurons $N_f$ begins to increase. This process can be divided into two stages: first the number $N_f$ increases slightly (first stage), and then very abruptly (second stage). The first stage takes part within a certain finite range of the coupling strength $\sigma$ and is related to partial synchronization (light green color in Fig.~\ref{fig:ring_ratio},(b,e)). This range is short in case of $\tau=1$, while it expands for a larger delay (Fig.~\ref{fig:ring_ratio},(a,b,d,e)). Moreover, for $\tau>2$, this $\sigma$-range is twice wider for a larger dissipation (Fig.~\ref{fig:ring_ratio},(a,b)) than in the case of smaller dissipation (Fig.~\ref{fig:ring_ratio},(d,e)). 
A single or several clusters of neurons with the oscillating synchronous dynamics are formed within this $\sigma$-range. The process of cluster formation is well illustrated by the spatial distributions of minimal values of the $x$ coordinate shown in Fig.~\ref{fig:ring_ratio},(c,f) for fixed $\tau=5$.  It is seen that the synchronous clusters are distributed randomly along the network space and have varying width. Our calculations have indicated (not shown in this paper) that this property depends on random initial condition distributions.    

The $\sigma$-range corresponding to the first stage is bounded by the second threshold value, $\sigma_{\rm th, all}$, where all the neurons are excited and $N_f=N$ (full synchronization in Fig.~\ref{fig:ring_ratio},(b,e)). Note that this transition occurs abruptly and the  second threshold is independent of $\tau$ starting with $\tau>1$ (Fig.~\ref{fig:ring_ratio},(a,d)). Similar as for the first threshold value described above, the value of  $\sigma_{\rm th, all}$ is larger for a larger dissipation (it is about $0.48$ for $\tau>1$ and $\gamma=0.5$,   Fig.~\ref{fig:ring_ratio},(a,b)) and smaller for a smaller dissipation (it is about $0.19$ for $\tau>1$ and $\gamma=0.7$, Fig.~\ref{fig:ring_ratio},(d,e)). 
Panels (c),(f) in Fig.~\ref{fig:ring_ratio} show that this two-stage process describes a first-order nonequilibrium phase transition from the quiescent state to full synchronization via partial cluster synchronization with increasing $\sigma$ for fixed $\tau=5$. This is similar to nucleation phenomena found in equilibrium and nonequilibrium systems, which have recently become a focus of research on synchronization of networks~\cite{fialkowski2023heterogeneous}. The final step is an abrupt single-step transition to full synchrony.
%However, there are not exact lines between the regions of global order and disorder in the network (Fig.~\ref{fig:ring_ratio},(b,e)), especially in the case of $\gamma=0.5$ (Fig.~\ref{fig:ring_ratio},(b)). Here we can observe an intermediate region for $\sigma\in [0.2,0.3]$ and $\tau>6$. Within this region almost all the neurons are quiescent and the order parameter $r$ is very close to 1. There is another intermediate region for $\sigma\in [0.45,0.6]$ and $\tau>6$, where, despite all the oscillators are firing, disorder appears due to the difference in the oscillatory phase of neighbouring neurons.

%%%%%%%%%%%%%%%%%%%%%%%%%%%%%%% fig.8c(14) %%%%%%%%%%%%%%%%%%%%%%%%%%%%%%%%%
\begin{figure}[!t]
\centering
\includegraphics[width=0.95\linewidth]{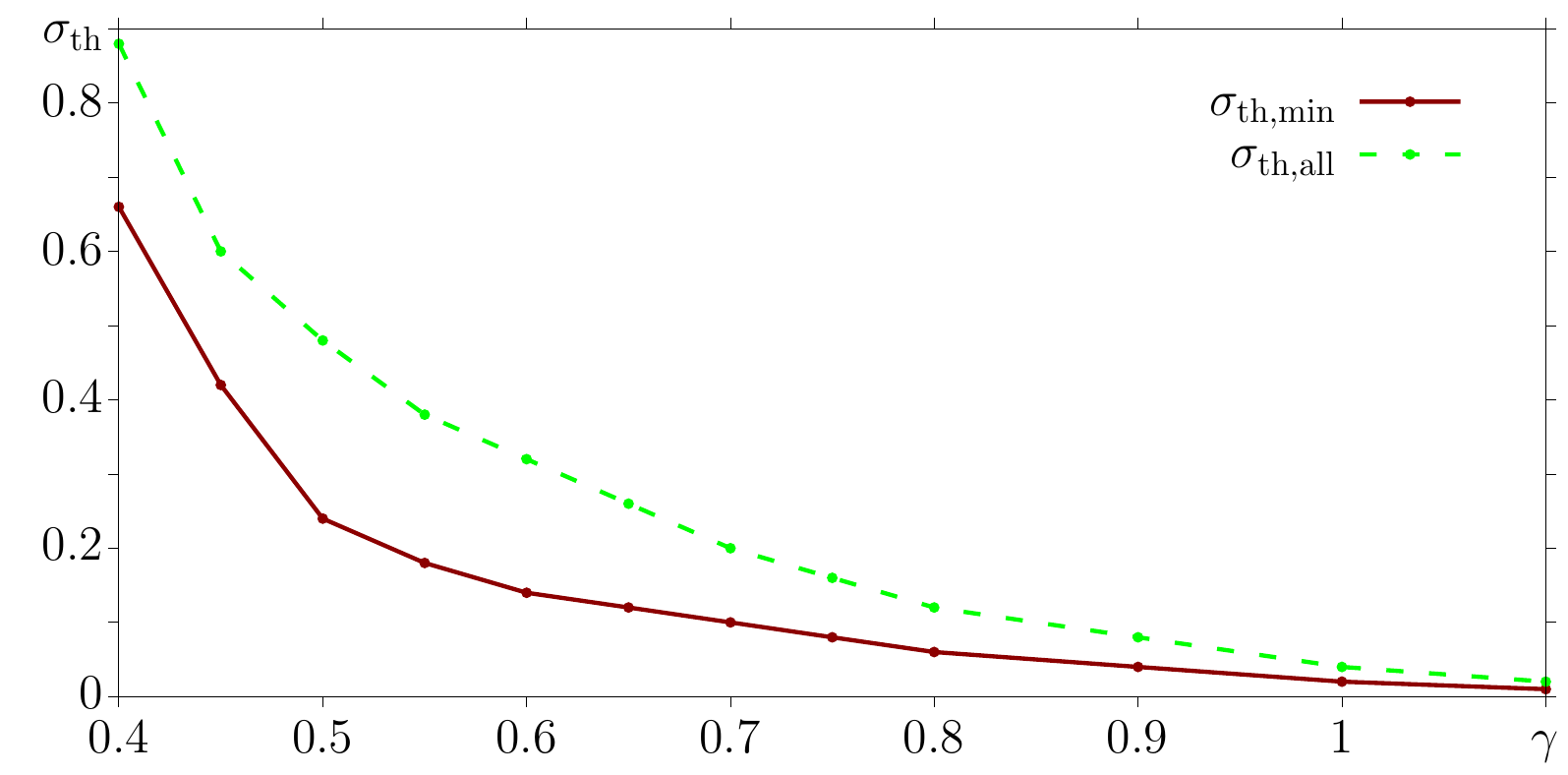}
\caption{Dependences of the threshold values $\sigma_{\text{th,min}}$ (red curve with dots) and $\sigma_{\text{th,all}}$ (green dashed curve with dots) versus the dissipation parameter $\gamma$ at $\tau=5$ for randomly distributed initial conditions in the network~\eqref{eq:ring}. Other parameters: $\varepsilon=0.01$, $\beta=-0.5$, $P=1$, $N=50$.}
\label{fig:ring_threshold}
\end{figure}
%%%%%%%%%%%%%%%%%%%%%%%%%%%%%%%%%%%%%%%%%%%%%%%%%%%%%%%%%%%%%%%%%%%%%%%%%

In order to get more insight into the effect of dissipation on the dynamics of FitzHugh-Nagumo neuron network, we calculate dependences of the two threshold values  $\sigma_{\text{th,min}}$ and $\sigma_{\text{th,all}}$ versus the dissipation parameter $\gamma$ when $\gamma$ is varied within the region of excitable dynamics (region III in Fig.~\ref{fig:FHN_modes}). The numerical results are plotted in Fig.~\ref{fig:ring_threshold}). It is seen that increasing $\gamma$ (decreasing the dissipation) leads to a gradual and nonlinear decrease in the thresholds of  oscillation excitation. Besides, these values are visibly different for $0.4<\gamma<0.9$ but when the dissipation parameter becomes very large, $\gamma>0.9$  (very low dissipation) and approaches the boundary of the self-sustained oscillatory regime, the threshold values almost coincide and vanish.  

\subsection{Linear stability analysis of the equilibrium}

The variational equation for a ring network of delay-coupled FitzHugh--Nagumo systems is given by 
\begin{equation}
\delta \dot{\mathbf{x}} = \frac{1}{\varepsilon} A \delta \mathbf{x}(t)  + \frac{\sigma}{\varepsilon} B \delta \mathbf{x}(t-\tau),
\label{eq:differentialCHainFHN}
\end{equation}
where $\mathbf{x} = (x_{1},y_{1},x_{2},y_{2},\ldots,x_{N},y_{N})$ and $N$ is the size of ring network considered. Here $N$ is set to be $50$.

The analysis is similar to the one performed for the simplified FitzHugh--Nagumo model\cite{Plotnikov2016}. For the network~\eqref{eq:differentialCHainFHN} we get the following expressions for the matrices $A$ and $B$:
\begin{align}
    A &= 
    \begin{bmatrix}
    \xi & -1 & 0 &0& \ldots &0&0\\
    \gamma \varepsilon & -\varepsilon & 0 & 0 & \ldots &0& 0\\
    0 & 0 & \xi & -1 & \ldots &0& 0\\
    0 & 0 & \gamma \varepsilon & -\varepsilon &\ldots &0&0\\
    \vdots &\ddots & \ldots & \vdots& & &\vdots\\
    0 & \ldots & 0 &0& \ldots & \xi &-1\\
    0 & \ldots & 0 &0& \ldots &\gamma \varepsilon & -\varepsilon
    \end{bmatrix},
    \label{eq:MatrixA}
\\
    B &= 
    \begin{bmatrix}
    0 & 0 & 1 & 0 & 0 & \ldots & 1 & 0\\
    0 & 0 & 0 & 0 & 0 & \ldots & 0 & 0\\
    1 & 0 & 0 & 0 & 1 & \ldots & 0 & 0\\
    0 & 0 & 0 & 0 & 0 & \ldots & 0 & 0\\
    \vdots &  &  &  & \ddots & \ldots & \vdots\\
    1 & 0 & \ldots &    &      1 &0 &0 &0\\
    0 & 0 & 0 & 0 & 0 & \ldots & 0 & 0\\
    \end{bmatrix},
\end{align}
where $\xi = 1 - {x_{1}^*}^2 - \sigma$. We apply the ansatz $\delta \mathbf{x}(t) = \exp(\lambda t) \mathbf{ u}$ to \eqref{eq:differentialCHainFHN}, which gives the following eigenvalue problem 
\begin{equation}
    \left(\frac{A}{\varepsilon} + \frac{\sigma B}{2\varepsilon} \exp(-\lambda \tau) \right) u = \lambda u.
    \label{eq:ChainEigenvalueProblem}
\end{equation}

%%%%%%%%%%%%%%%%%%%%%%%%%%%%%%% fig.11 %%%%%%%%%%%%%%%%%%%%%%%%%%%%%%%%%
\begin{figure}[!t]
\centering
$\gamma=0.5$ \hspace{93pt} $\gamma=0.7$
\\
\includegraphics[width=0.495\linewidth]{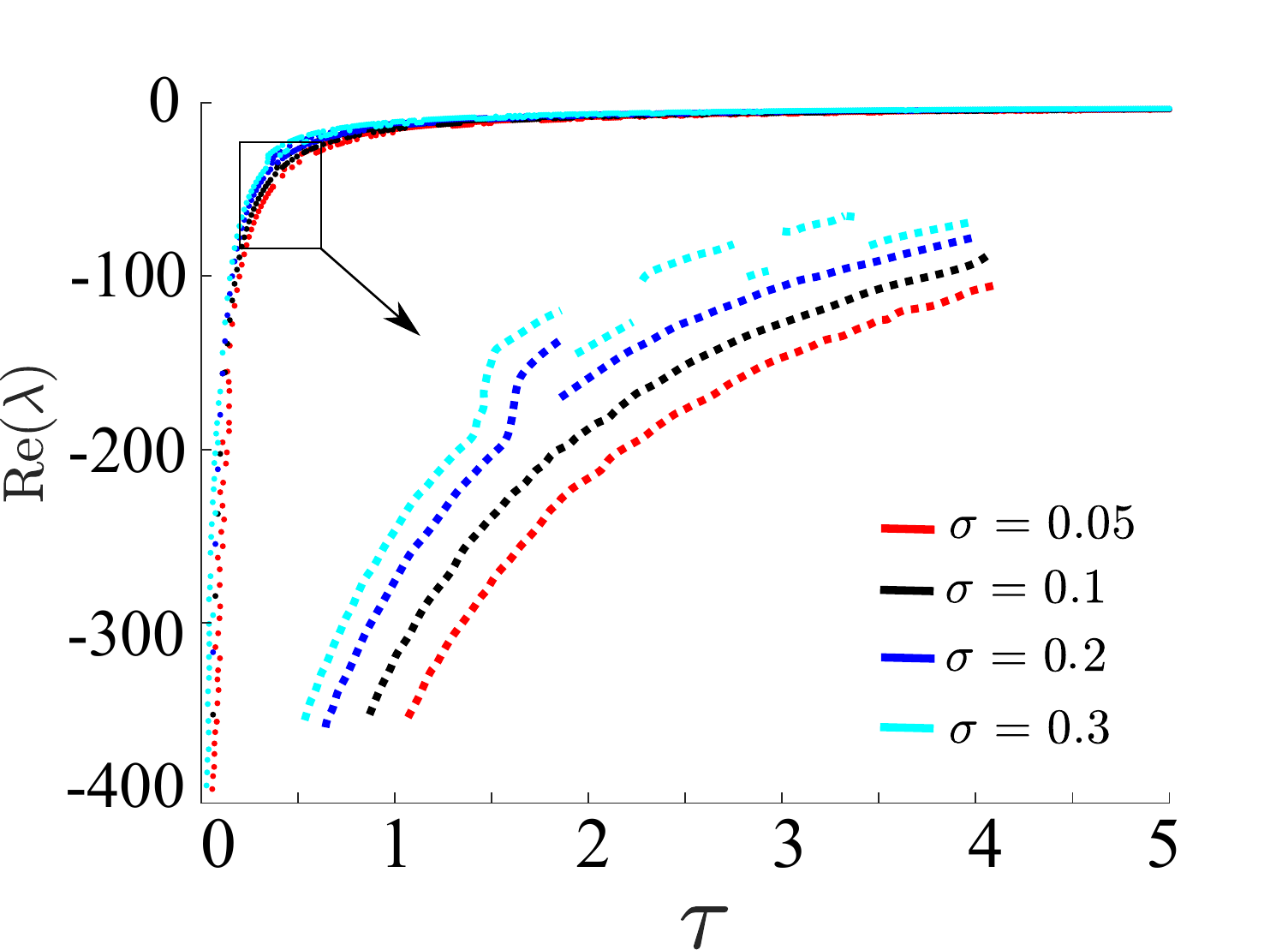}
\includegraphics[width=0.495\linewidth]{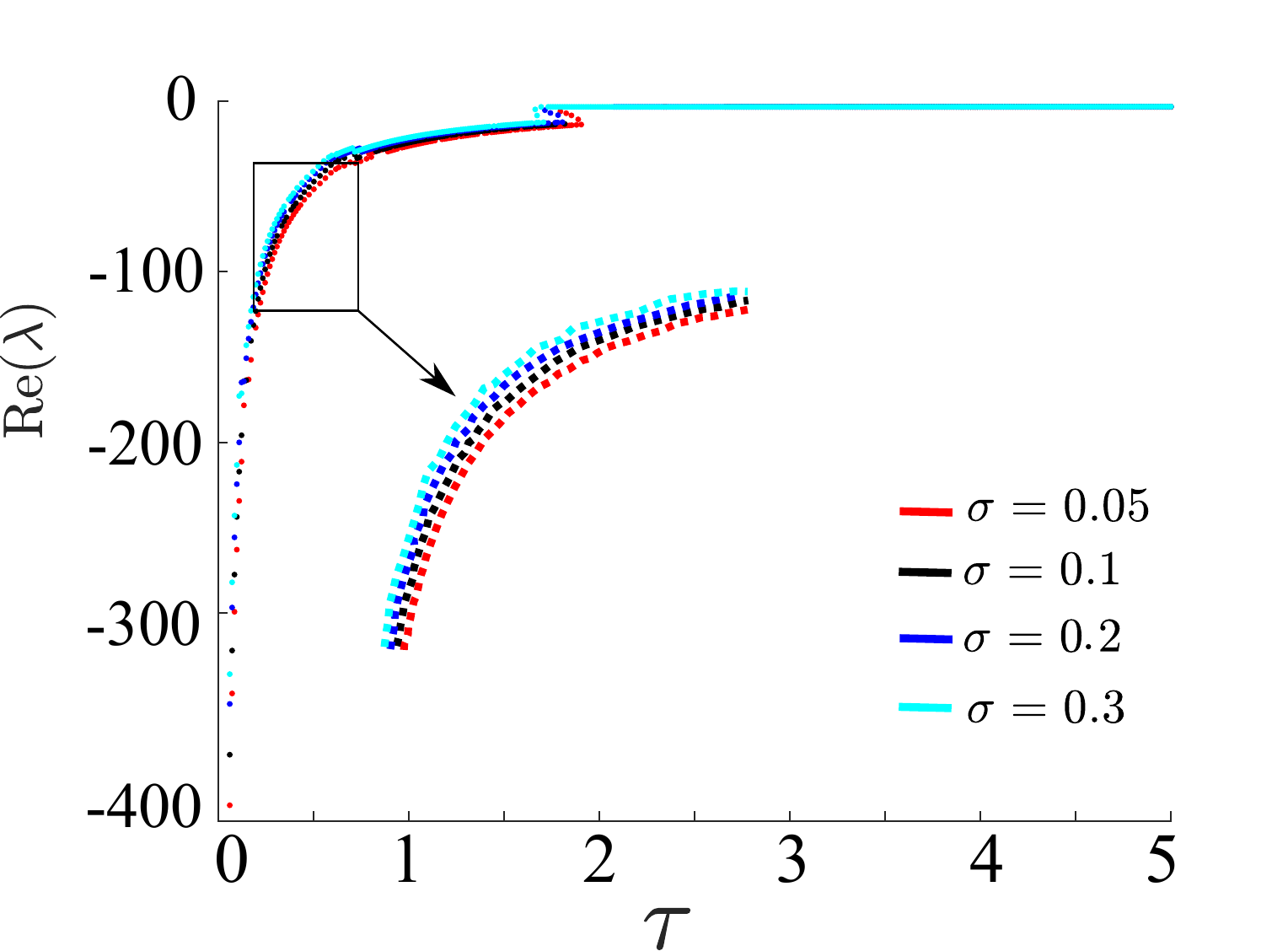}
\\
(a)\hspace{113pt}(b)
\caption{Real part of the eigenvalues $\Lambda$ vs delay time $\tau$ for  $\gamma=0.5$ (a) and  $\gamma=0.7$ (b). The equilibrium point of system~\eqref{eq:differentialCHainFHN} for the ring of coupled FitzHugh--Nagumo units is stable for all parameter values shown. The curve shifts slightly upwards with increasing $\sigma$ (see blow-up in the insets). The parameters are set as $\varepsilon=0.01, \beta=-0.5$.}
\label{fig:NCoupledStabilityAnalysis}
\end{figure}
%%%%%%%%%%%%%%%%%%%%%%%%%%%%%%%%%%%%%%%%%%%%%%%%%%%%%%%%%%%%%%%%%%%%%%%%%

To obtain the values of $\lambda$, we solve the determinant
$$\Bigg|\frac{A}{\varepsilon} + \frac{\sigma B}{2\varepsilon} e^{-\lambda \tau} - \lambda I_{N \times N}\Bigg| = 0.$$
We also observe a similar trend in the $\rm{Re}(\lambda)$ versus $\tau$ plot for the ring of FitzHigh-Nagumo units as in Sec.~\ref{secIIIB}. For all values of $\tau$, the real part is negative and the modulus of the real part of the eigenvalue decreases as $\tau$ increases and then tends to a fixed negative value for larger $\tau$. Hence the equilibrium point of the system is stable. With an increase in $\sigma$ as shown in different colours in Fig.~\ref{fig:NCoupledStabilityAnalysis}, the curve slightly shifts upwards. With an increase in $\gamma$ to $0.7$, we see that the shape of the curve is similar. In conclusion, the quiescent equilibrium is always stable. Therefore, the equilibrium point coexists with limit cycle oscillations in regions where the oscillatory regime is observed, and there is multistability.

\section{Influence of nonlocal coupling on delay-induced oscillations}
 
\begin{figure}[!t]
\centering
\includegraphics[width=0.495\linewidth]{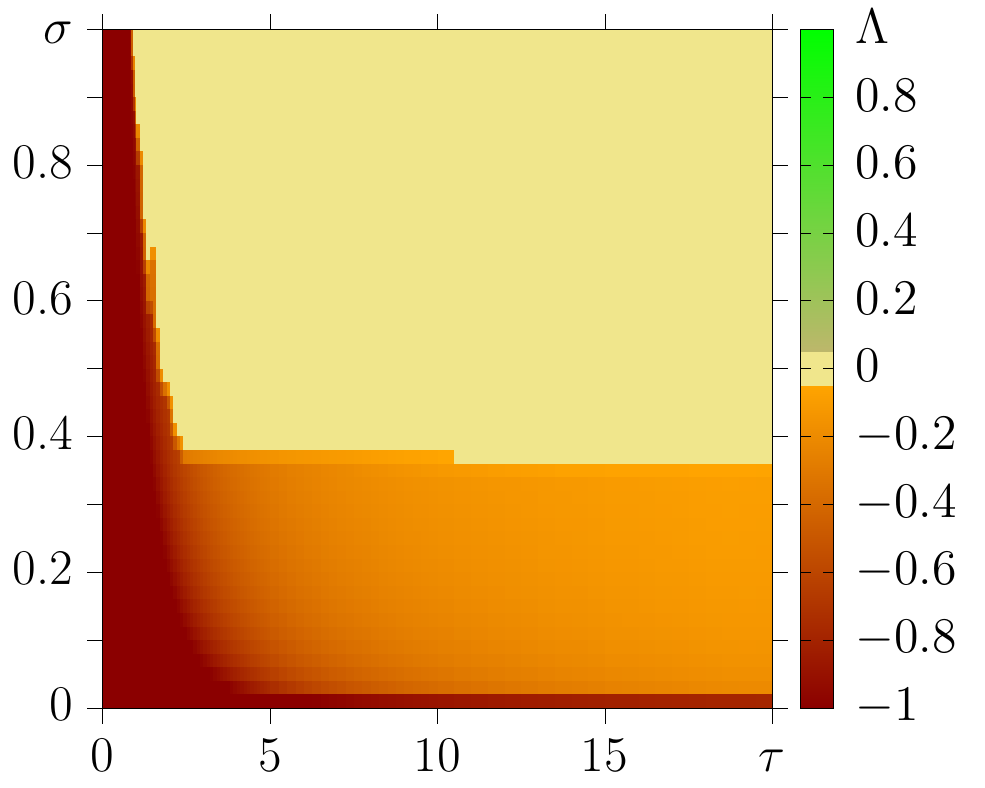}
\includegraphics[width=0.495\linewidth]{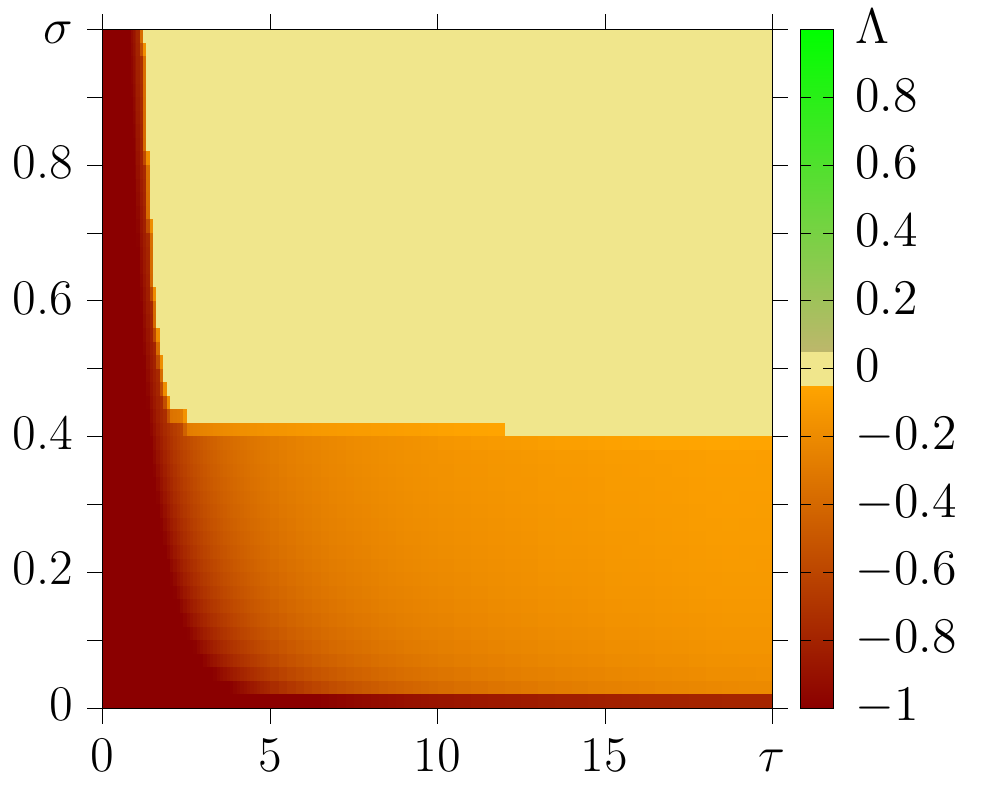}
\\
(a)\hspace{113pt}(b)
\\
\includegraphics[width=0.495\linewidth]{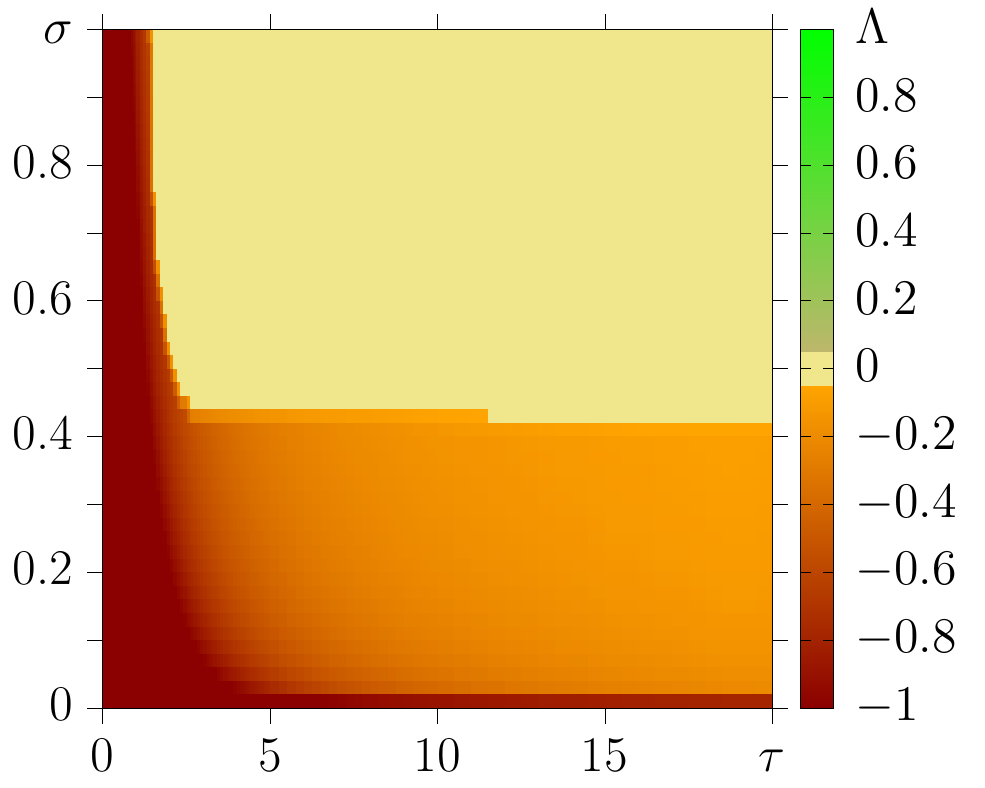}
\includegraphics[width=0.495\linewidth]{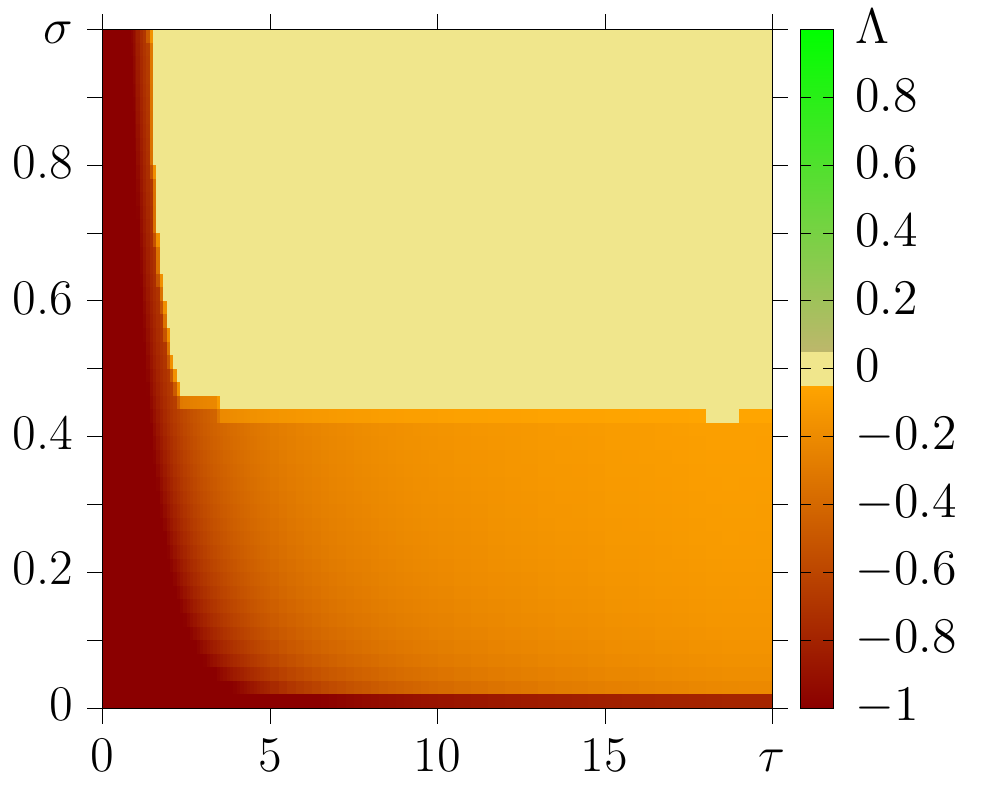}
\\
(c)\hspace{113pt}(d)
\caption{Distributions the maximal Lyapunov exponent $\Lambda$ in the ($\tau,\sigma$) parameter plane for the network~\eqref{eq:ring}  for four different values of the coupling range $P$: (a) $P=2$, (b) $P=3$, (c) $P=4$, and (d) $P=5$. Other parameters:  $\varepsilon=0.01$, $\beta=-0.5$, $\gamma=0.5$, $N=50$.}
\label{fig:ring_nonlocal}
\end{figure}

To determine the effects of nonlocal coupling, we study the oscillations in the oscillatory ring network~\eqref{eq:ring} when each neuron interacts with $P$ neighbors from the left and from the right. We vary the coupling range within  $P \in [2,5]$ and calculate distributions of the maximal Lyapunov exponent $\Lambda$ in the ($\tau,\sigma$) parameter plane within the same intervals of $\tau$ and $\sigma$ variation as in Fig.~\ref{fig:ring_regimes}. Numerical results are shown in Fig.~\ref{fig:ring_nonlocal} for the fixed value of $\gamma=0.5$. If we compare these diagrams with those for the local coupling (Fig.~\ref{fig:ring_regimes},(a)), we can see that introducing nonlocal coupling leads to a significant increase in the threshold values for inducing oscillations with respect to both the delay time $\tau$ and the coupling strength $\sigma$. Interestingly, there is a large difference between the threshold level for the cases $P=1$ (Fig.~\ref{fig:ring_regimes},(a)) and $P=2$ (Fig.~\ref{fig:ring_nonlocal},(a)). However, a further growth of the coupling range $P$ leads only to a weak increase in the thresholds with respect to $\sigma$ and $\tau$.
%%%%%%%%%%%%%%%%%%%%%%%%%%%%%%% fig.12 %%%%%%%%%%%%%%%%%%%%%%%%%%%%%%%%%
%%%%%%%%%%%%%%%%%%%%%%%%%%%%%%%%%%%%%%%%%%%%%%%%%%%%%%%%%%%%%%%%%%%%%%%%%

\begin{figure}[!t]
\centering
\includegraphics[width=0.495\linewidth]{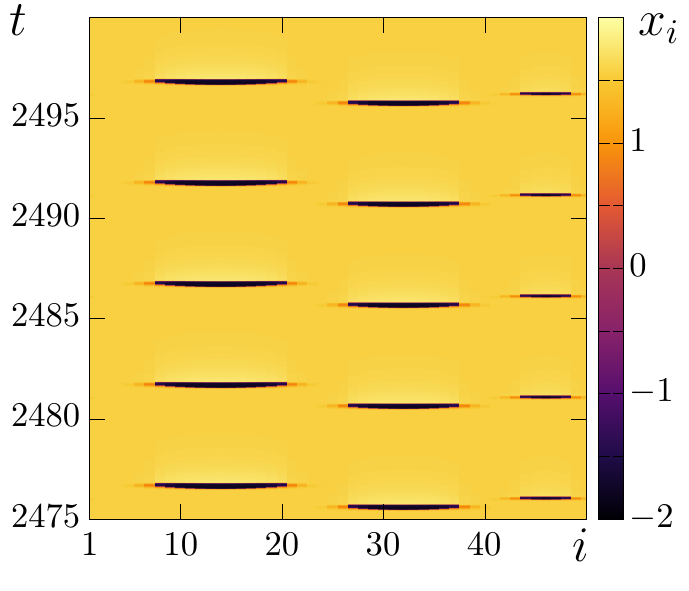}
\includegraphics[width=0.495\linewidth]{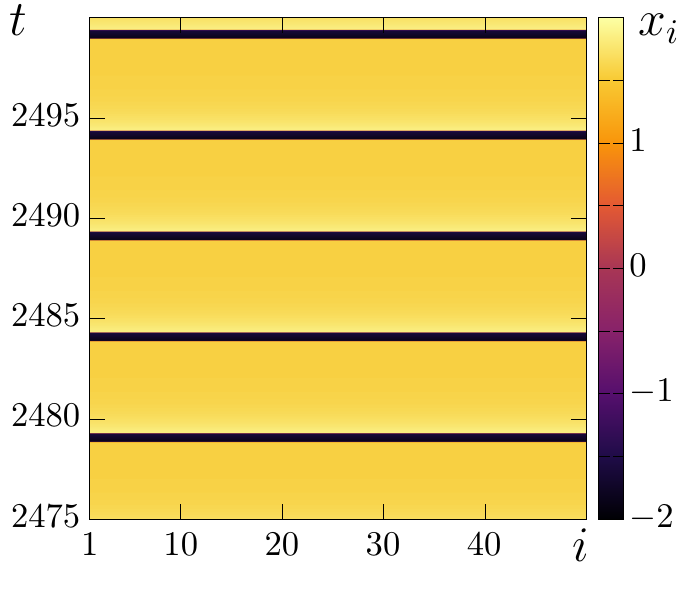}
\\
(a)\hspace{113pt}(b)
\\
\includegraphics[width=0.495\linewidth]{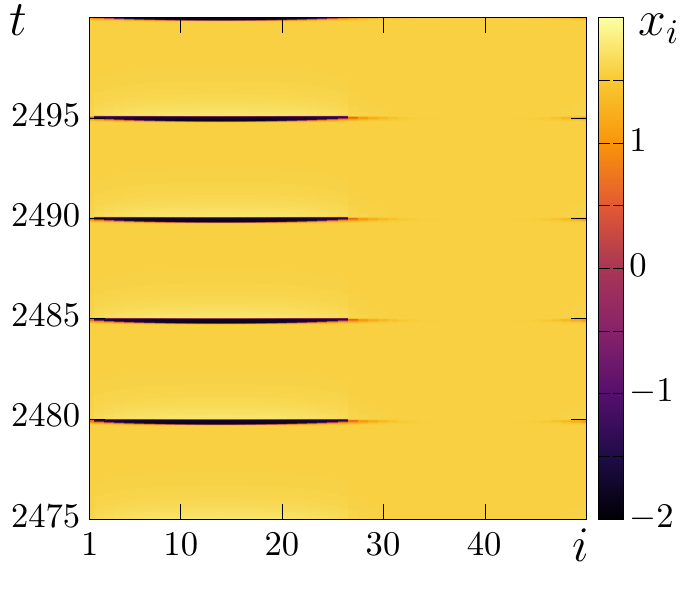}
\includegraphics[width=0.495\linewidth]{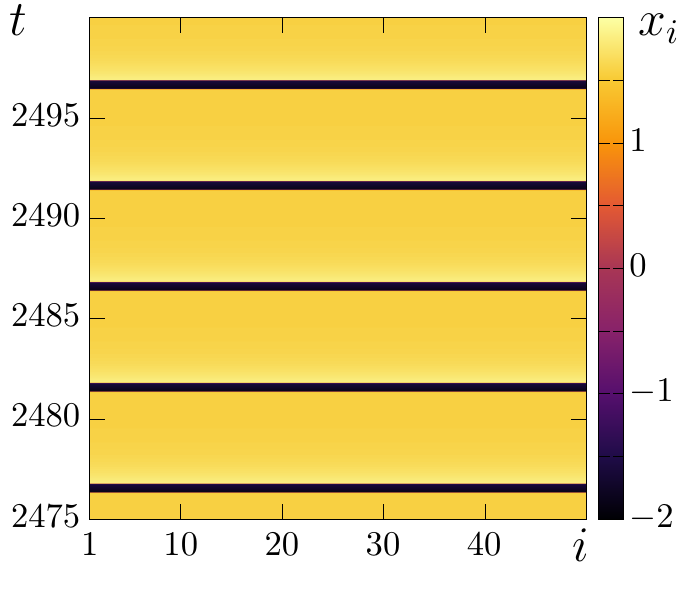}
\\
(c)\hspace{113pt}(d)
\caption{Delay-induced oscillations in case of nonlocal coupling shown in asymptotic spatio-temporal diagrams for a system~\eqref{eq:ring} at (a) $P=2,\sigma=0.42$; (b) $P=2,\sigma=0.48 $; (c) $P=5,\sigma=0.46$; and (d) $P=5,\sigma=0.48$. Other parameters: $\varepsilon=0.01$, $\beta=-0.5$, $N=50$.}
\label{fig:ring_nonlocal_regimes}
\end{figure}

At the same time, the dynamical regimes do not undergo any qualitative changes in comparison with the case of local coupling $P=1$ illustrated in Fig.~\ref{fig:ring_STP}. The pulse duration and spatial features remain very similar. The spatiotemporal dynamics of these regimes are illustrated by space-time plots for $P=2$ and $P=5$ in Fig.~\ref{fig:ring_nonlocal_regimes},(a,b) and ~\ref{fig:ring_nonlocal_regimes},(c,d), respectively. The left column in Fig.~\ref{fig:ring_nonlocal_regimes} demonstrates the case when only a part of the ring is firing (lower values of the coupling strength), while the right column corresponds to the case when the whole system fires. Thus, the first case is similar to that one shown in Fig.~\ref{fig:ring_STP},(b2) and the second column  looks like the regime in Fig.~\ref{fig:ring_STP},(c2) for the system with local coupling $P=1$. As was shown\cite{Zakharova2020}, it is typical for nonlocal coupling that multiple clusters are found for small coupling range, and single clusters for larger coupling range.

%%%%%%%%%%%%%%%%%%%%%%%%%%%%%%% fig.13 %%%%%%%%%%%%%%%%%%%%%%%%%%%%%%%%%
%%%%%%%%%%%%%%%%%%%%%%%%%%%%%%%%%%%%%%%%%%%%%%%%%%%%%%%%%%%%%%%%%%%%%%%%%

\section{Conclusions}

Our detailed numerical analysis of the dynamics of the networks of delay-coupled FitzHugh--Nagumo oscillators with dissipation has shown qualitative and in some cases quantitative similarities with the dynamics of delay-coupled simplified FitzHugh--Nagumo models\cite{Dahlem2009, Scholl2009, Panchuk2013, Nikitin2019}. However, a number of specific differences have been established, which are due to the dissipation parameter $\gamma$ in the considered neuron model. 

Our numerical simulations of the dynamics of two delay-coupled FutzHugh-Nagumo oscillators have demonstrated that the threshold of inducing both in-phase and anti-phase  periodic oscillations in the system  with respect to the coupling strength $\sigma$ substantially depends on the dissipation parameter $\gamma$ when the delay time $\tau$ increases. The larger $\gamma$ and thus the smaller the dissipation in the neuron, the less the $\sigma$ threshold. The smaller dissipation means that a substantially lower amplitude of excitation is needed to induce oscillations in the FitzHugh--Nagumo neurons. Starting with a certain value of $\tau$, the $\sigma$ threshold remains unchanged as the delay time increases. The threshold for the oscillatory regime with respect to the delay time $\tau$ does not depend on the dissipation parameter and, as our calculations have shown, it is larger for inducing in-phase oscillations than that for anti-phase oscillations. 
The same peculiarity in the $\sigma$ threshold  is observed for the ring network of locally delay-coupled FitzHugh--Nagumo oscillators. However, in this case the dissipation parameter $\gamma$ also has a significant effect on the $\tau$ threshold value for the occurrence of oscillations. When the dissipation is low ($\gamma$ is large), the delay-induced oscillations emerge in the network for smaller values of $\tau$ as compared with the case of large dissipation. A sufficiently high level of dissipation prevents neurons to switch to the self-oscillatory mode, i.e. stronger coupling is needed to excite neurons.

We have analyzed in detail the transition of locally delay-coupled FitzHugh--Nagumo neurons from the quiescent to the oscillatory state for two different values of the dissipation parameter $\gamma$ when the coupling strength $\sigma$ and the delay time $\tau$ are varied. For this purpose we have calculated and constructed the distributions of the global order parameter and the ratio of the firing neurons to the whole number of the network nodes. Both two-parameter diagrams have indicated the presence of two different threshold values with respect to $\sigma$. The first value corresponds to the case when only a small part of the neurons are excited, i.e., synchronous clusters are formed, and the second one -- when all the neurons demonstrate self-sustained oscillations, i.e. full synchrony is reached. It has been established that both threshold values of $\sigma$ are substantially smaller (at least twice) for weaker dissipation than for stronger dissipation. The intermediate region with respect to $\sigma$ (starting with a small delay time $\tau \approx 2$) where the network neurons are consequently excited is three times narrower for large $\gamma$ as compared with the case of smaller $\gamma$. With this, we have established a first-order nonequilibrium phase transition from the quiescent state to the fully synchronous state via cluster formation with increasing coupling strength. The second, abrupt transition from clusters to full synchrony is a single-step transition where multiple clusters abruptly merge into one at a certain threshold coupling strength; it is similar to nucleation phenomena recently observed in transitions to synchrony in power grids~\cite{tumash2018effect} and adaptive neuronal networks~\cite{fialkowski2023heterogeneous}.
 
We believe that the results presented in this work for networks of delay-coupled FitzHugh--Nagumo neurons with dissipation will contribute to the insight into synchronization transitions by generalizing previous results obtained for the simplified FitzHugh--Nagumo networks.

\begin{acknowledgments}

The reported study was funded by the Russian Science Foundation (project No. 20-12-00119). E.S. acknowledges financial support from the German Science Foundation (DFG-Projektnummer 163436311--SFB 910, 429685422 and 440145547). A. V. Bukh thanks for the financial support provided by The Council for grants of President of Russian Federation, project number SP-774.2022.5.

\end{acknowledgments}

\section*{Data Availability Statement}

The data that support the findings of this study are available
from the corresponding author upon reasonable request.

%\appendix

%\section{Appendixes}

%To start the appendixes, use the \verb+\appendix+ command.

%\section{A little more on appendixes}

%\subsection{\label{app:subsec}A subsection in an appendix}

%\nocite{*}
\bibliography{mybibfile}% Produces the bibliography via BibTeX.

\end{document}